# A review of the ocean-atmosphere interactions during tropical cyclones in the north Indian Ocean

**Vineet Kumar Singh**[1,2] and **M.K. Roxy**[1*]

[1] *Indian Institute of Tropical Meteorology, Ministry of Earth Sciences, Pune, India*

[2] *Department of Atmospheric and Space Sciences, Savitribai Phule Pune University, Pune, India*

*Corresponding author address: Roxy Mathew Koll, Indian Institute of Tropical Meteorology, Pune 411008, India. E-mail: roxy@tropmet.res.in

**Abstract**

The north Indian Ocean accounts for 6% of the global tropical cyclones annually. Despite the small fraction of cyclones, some of the most devastating cyclones have formed in this basin, causing extensive damage to the life and property in the north Indian Ocean rim countries. In this review article, we highlight the advancement in research in terms of ocean-atmosphere interaction during cyclones in the north Indian Ocean and identify the gap areas where our understanding is still lacking.

There is a two-way ocean-atmosphere interaction during cyclones in the north Indian Ocean. High sea surface temperatures (SSTs) of magnitude 28–29°C and above provide favorable conditions for the genesis and evolution of cyclones in the Arabian Sea and the Bay of Bengal. On the other hand, cyclones induce cold and salty wakes. Cyclone induced cooling depends on the translation speed of the cyclone, wind power input, ocean stratification, and the subsurface conditions dictated by the ocean eddies, mixed layer and the barrier layer in the north Indian Ocean. The average cyclone-induced SST cooling is 2–3°C during the pre-monsoon season and 0.5–1°C during the post-monsoon season. This varying ocean response to cyclones in the two seasons in the Bay of Bengal is due to the difference in the ocean haline stratification, whereas, in the Arabian Sea it is due to the difference in cyclone wind power input and ocean thermal stratification. The oceanic response to cyclone is asymmetric due to the asymmetry in the cyclone wind stress, cyclone induced rainfall and the dynamics of the ocean inertial currents. The cyclone induced wake is salty and is the saltiest in the Bay of Bengal among all the ocean basins. The physical response of the ocean to the cyclone is accompanied by a biological response also, as cyclones induce large chlorophyll blooms in the north Indian Ocean that last from several days to weeks.

SSTs leading to cyclogenesis in the Arabian Sea are 1.2–1.4°C higher in the recent decades, compared to SSTs four decades ago. Rapid warming in the north Indian Ocean, associated with global warming, tends to enhance the heat flux from the ocean to the atmosphere and favor rapid intensification of cyclones. Monitoring and forecasting rapid intensification is a challenge, particularly due to gaps in in-situ ocean observations. Changes in ocean-cyclone interactions are emerging in recent decades in response to Indian Ocean warming, and are to be closely monitored with improved observations since future climate projections demonstrate continued warming of the Indian Ocean at a rapid pace along with an increase in the intensity of cyclones in this basin.

# 1. Introduction

## 1.1. Overview of ocean-cyclone interactions

Tropical cyclones genesis and intensification occur mainly over the regions where the sea surface temperatures (SSTs) are greater than 26ºC, mid-tropospheric humidity is high, and vertical wind shear is low (Gray 1968). The ocean plays a dominant role in providing energy to the tropical cyclone through the exchange of latent and sensible heat fluxes between the ocean and the atmosphere (Ooyama 1969; Emanuel 1986; Vinod et al. 2014). Increased latent heat flux from the ocean to the atmosphere moistens the boundary layer and provides conducive conditions for cyclogenesis (Gao et al. 2019) and also plays a crucial role in its intensification (Gao and Chiu 2010; Lin et al. 2014; Jaimes et al. 2015; Gao et al. 2016). There is a two-way interaction between SSTs and tropical cyclones (Emanuel 1986). SSTs significantly modulate the cyclone intensity (Kaplan and DeMaria 2003). High SSTs, favor an increase in the intensity of tropical cyclones through heat flux transfer from the ocean to the atmosphere (Emanuel 1986). On the other hand, the cyclone passage across the ocean leads to a decrease in the SSTs, which in turn reduces the cyclone intensity due to a reduction in the transfer of heat flux from the ocean to the atmosphere (DeMaria and Kaplan 1999; Cione and Uhlhorn 2003; Lloyd and Vecchi 2011). Cyclone induced cooling not only affects the intensity of cyclone but also has a pronounced effect on the cyclone size (Chen et al. 2010; Ma et al. 2013) and structure (Zhu et al. 2004; Chen et al. 2010). The cold wake induced by the cyclone reduces the strength of the warm core and the secondary circulation of the cyclone gets weakened. The outward expansion of the cyclonic circulation is also hindered, which leads to a decrease in the cyclone size (Ma et al. 2013). Cyclone cools the ocean by multiple processes, such as friction-induced vertical mixing due to the high wind speed of cyclone, upwelling of cool subsurface water to the ocean surface, internal waves in the ocean due to tropical cyclone winds (Brooks 1983; Shay and Elsberry 1987) and air-sea heat fluxes (Shen and Ginis 2003). Earlier studies for different ocean basins have shown that out of the different processes associated with the cyclone induced cooling, the vertical mixing associated with the upwelling dominates the cyclone cold-wake, with a share of 70-80% (Price 1981; Jullien et al. 2012; Vincent et al. 2012a). On the other hand, the cooling due to horizontal advection and enthalpy flux contributes only about 20% (Shay et al. 1992; Huang et al. 2009; Vincent et al. 2012a). For weak cyclones, the contribution of the enthalpy flux in the cooling is larger than the vertical mixing (Vincent et al. 2012a).

The effect of the ocean-cyclone interaction on the cyclone track is yet to be clearly understood. A few studies highlight that ocean-cyclone interaction improve the realistic representation of the cyclone track in the model and improve the climatological representation of cyclone tracks (Subrahmanyam et al. 2005; Jullien et al. 2014; Ogata et al. 2015). Whereas other studies show that the tracks of the cyclones are generally not affected by the ocean-cyclone coupling (Zhu et al. 2004; Chen et al. 2010; Liu et al. 2011).

1.2. Characteristics of the north Indian Ocean controlling the ocean-cyclone interactions

As seen from the earlier discussion, multiple factors control the ocean-cyclone interaction which is dependent on cyclone and ocean characteristics. The north Indian Ocean sees significant seasonal variations in the SST, salinity, the mixed layer and the barrier layer (Figure 1). In the pre-monsoon season, the SSTs throughout the north Indian Ocean are high of the order of 29-30°C (Figure 1a). Whereas, in the post-monsoon season, the SSTs are roughly 1°C cooler than the pre-monsoon season (Figure 1b). Similarly, salinity also has a seasonality variation especially in the Bay of Bengal (Figure 1c-d). In the post-monsoon season, the Bay of Bengal has low salinity in the northeast part and high salinity in the southwest part of the basin (Rao and Sivakumar 2003). The low salinity in the northern part of the Bay of Bengal in this season is mainly due to the oceanic rainfall and the huge river discharge into the northern Bay of Bengal (Shetye et al. 1996; Shenoi 2002; Sengupta et al. 2006; Akhil et al. 2014; Chaitanya et al. 2014). It leads to the formation of a very strong near-surface halocline along the rim of the Bay of Bengal during the post-monsoon season. Due to the presence of freshwater at the surface, the mixed layer becomes shallower (Figure 1f). This enhances the ocean stratification with freshwater at the surface and saline water in the subsurface (Vinayachandran et al. 2002) leading to the formation of a strong barrier layer (Figure 1h) (Thadathil et al. 2007). This seasonal variation of barrier layer is absent in the Arabian Sea. However, the mixed layer depth in the Arabian Sea shows large seasonal variations as compared to the Bay of Bengal (Figure 1e-1f) (Shenoi 2002; Prasad 2004; Babu et al. 2004; Narvekar and Prasanna Kumar 2006). The lack of seasonal variability in the mixed layer is mainly observed in the north Bay of Bengal due to haline stratification. In the south Bay of Bengal the mixed layer depth shows seasonal variability with a deep mixed layer in the monsoon and winter season and a shallow mixed layer in the spring and autumn season. All these ocean characteristics have a pronounced impact on the ocean-cyclone interaction as will be discussed later.

## 1.3. Statistics of cyclone in the north Indian Ocean

The north Indian Ocean, including the Arabian Sea and the Bay of Bengal, has two cyclone seasons in a year, during April–early June (pre-monsoon season), and October–December (post-monsoon season). The pre-monsoon and post-monsoon cyclone seasons are separated by the monsoon season during June–September. Cyclones, generally do not form in the monsoon season due to the strong wind shear prevailing in this season (Li et al. 2013). The north Indian Ocean accounts for 6% of the tropical cyclones globally (Figure 2, Neumann 1993; Ali 1999). Despite the relatively smaller number of cyclones, more than 80% of global fatalities occur due to tropical cyclones in this region, mostly due to coastal flooding (Beal et al. 2020). Several of the most devastating cyclones that formed in the Bay of Bengal caused extensive casualties in the Bay of Bengal rim countries, particularly in India, Bangladesh, and Myanmar (Madsen and Jakobsen 2004; Kikuchi et al. 2009; McPhaden et al. 2009). Basin-wise, the Bay of Bengal cyclone contribution is 4%, and the Arabian Sea cyclone contribution is 2% to the annual global frequency of the cyclones (Figure 2). These statistics based on the recent data (1980–2019) is similar to the earlier reported statistics (Alam et al. 2003). High SST, higher moisture content over the Bay of Bengal and the propagation of disturbances from the northwest Pacific Ocean into the Bay of Bengal leads to a higher frequency of cyclones in this basin as compared to the Arabian Sea (Sahoo and Bhaskaran 2016; Singh et al. 2019).

Figure 3 shows cyclone tracks in the pre-monsoon and the post-monsoon season in the Arabian Sea and the Bay of Bengal during the period 1980–2019. During the period 1980–2019, in the pre-monsoon season, 26% of the north Indian Ocean cyclones intensify to category 3 or higher intensity cyclone. However, in the post-monsoon season only 15% of the north Indian Ocean cyclones intensify to category 3 or higher intensity cyclone. In this period, the strongest cyclone in the Arabian Sea is Cyclone Gonu in June 2007, with a maximum wind speed of 145 knots. Whereas, the strongest cyclone in the Bay of Bengal is Cyclone Fani in April 2019, with a maximum wind speed of 150 knots. It can be seen that in the Bay of Bengal, out of the two cyclone seasons, more cyclones form in the post-monsoon season, which is mainly because of the high relative humidity in this season (Li et al. 2013).

1.4. Cyclone interactions with subseasonal (MJO) and interannual variability (ENSO and IOD)

Along with the local interaction of cyclones with the ocean, the frequency and intensification of the cyclones in the north Indian Ocean are also influenced by the Madden-Julian Oscillation (MJO) at the subseasonal scale (Krishnamohan et al. 2012; Singh et al. 2020). MJO is characterized by an eastward propagating band of enhanced convection in the tropical belt close to the equator and has a periodicity of 30-60 days (Madden and Julian 1971, Roxy et al. 2019). The cyclone genesis in the north Indian Ocean gets clustered in the region of enhanced low-level relative vorticity, low to mid-level relative humidity (Singh et al. 2021) and reduced wind shear (Tsuboi and Takemi 2014) associated with the MJO (Liebmann et al. 1994; Roman-Stork and Subrahmanyam 2020).

The cyclone activity in the north Indian Ocean is also affected at the interannual timescale by the changes in the large-scale ocean-atmosphere conditions such as El Niño Southern Oscillation (ENSO). An increase in cyclone frequency, intensity and higher rate of rapid intensification of cyclone is observed especially in the Bay of Bengal during La Niña years as compared to the El Niño years (Felton et al. 2013; Girishkumar et al. 2015; Mahala et al. 2015; Bhardwaj et al. 2019). This is due to favorable atmospheric conditions such as anomalous high relative humidity, high relative vorticity and anomalous low wind shear in the Bay of Bengal during the La Niña years (Ng and Chan 2012; Felton et al. 2013; Bhardwaj et al. 2019).

Similar to ENSO, Indian Ocean Dipole (IOD) also has a significant effect on the cyclone frequency in the north Indian Ocean (Yuan and Cao 2013; Li et al. 2016). A positive IOD is characterized by anomalous cooler SSTs in the eastern tropical Indian Ocean and anomalous warmer SSTs in the western tropical Indian Ocean (Saji et al. 1999). Whereas, a negative IOD is characterized by anomalous warmer SSTs in the eastern tropical Indian Ocean and anomalous cooler SSTs in the western tropical Indian Ocean (Saji et al. 1999). During negative IOD years, anomalous warm SSTs and favorable atmospheric conditions such as enhanced relative vorticity and relative humidity in the eastern tropical Indian Ocean coupled together provide favorable conditions for cyclone formation in the Bay of Bengal (Mahala et al. 2015; Li et al. 2016).

The key objective of the current review is to provide an overview of the scientific progress in our understanding of ocean-cyclone interactions in the north Indian Ocean and to identify the gap areas where our understanding is still lacking. Section 2 details the various indices used to study cyclone characteristics and the methodology used. Section 3 discusses the role of the ocean in modulating cyclone characteristics and section 4 discusses the ocean dynamics in response to the cyclones. Section 5 assess our understanding of ocean biological response to cyclones. In section 6, we highlight the coupled interactions between the north Indian Ocean cyclones and the climate system. Section 7 discusses the observed and projected changes in ocean-cyclone interactions and section 8 summarizes and the key points along with the knowledge gap area in cyclone-ocean interactions.

## 2. Data, methodology and cyclone metrics

The review on ocean-cyclone interaction is supplemented by extending the analysis with cyclone data from the International Best Track Archive for Climate Stewardship (IBTrACS) cyclone dataset (Knapp et al. 2010), for the period 1980–2019. In order to elucidate the role of SSTs in the ocean-cyclone interaction, the daily SST data for the period 1982–2019, at a spatial resolution of 0.25° is obtained from Optimum Interpolation Sea Surface Temperature (OISST) dataset provided by the National Oceanic and Atmospheric Administration (NOAA) (Reynolds et al. 2007). The SST prior to the genesis of the cyclone is the SSTs averaged for a week (day -7 to day -1) prior to the day of cyclogenesis (day 0), over a 5°x5° region around the genesis center. Composite change in SST due to cyclones in the north Indian Ocean during the period 1982–2019, is estimated as the difference between SST averaged from day +1 to day +5 after the passage of the cyclone passage and the SST averaged from day -6 to day -2 before the passage of the cyclone, over a 1°x1° region where and when the cyclone attained its maximum wind speed (referred to as day 0). To show the cyclone category wise influence on the SSTs the cyclone category based on the maximum wind speed is defined as: category 1 (119–153 km h$^{-1}$), category 2 (154–177 km h$^{-1}$), category 3 (178–208 km h$^{-1}$), category 4 (209–251 km h$^{-1}$) and category 5 ($\geq$ 252 km h$^{-1}$). Tropical storms (JTWC definition, 63–118 km h$^{-1}$) that are considered as cyclones by India Meteorological Department (IMD) are included in Category 0.  Rapid intensification of the cyclone is defined as the increase in the wind speed of the cyclone by at least 30 knots ($\geq$ 55.6 km h$^{-1}$) in 24 hours (Kaplan and DeMaria 2003).

Climatological SST and salinity from the World Ocean Atlas, mixed layer depth and barrier layer thickness from de Boyer Montégut et al. (2007) are used to show the background ocean conditions during the pre- and post-monsoon seasons.

2.1. Genesis potential parameter index

The genesis potential parameter is used (GPP) to monitor and forecast cyclogenesis in the north Indian Ocean,  based on dynamic and thermodynamic atmospheric conditions (Kotal et al. 2009, equation 1).

GPP = ((ξ850 x M x I)/S)          if ξ850 >0, M >0 and I >0      Equation 1

     =   0                      if ξ850 ≤0, M ≤0 or I ≤0

Where $\xi 850$ = low-level relative vorticity at 850 hPa, M is the middle troposphere relative humidity given by the formula M = (RH-40)/30. In this, RH is the average relative humidity between 700 hPa and 500 hPa, I is the mid-troposphere instability which is the temperature difference (°C) between 850 hPa and 500 hPa. S is the vertical wind shear of horizontal winds between 200 hPa and 850 hPa (m s$^{-1}$).

This GPP index does not consider the ocean conditions that are important for cyclogenesis. To include the influence of the ocean in cyclogenesis, Suneeta and Sadhuram (2018) included Tropical Cyclone Heat Potential (TCHP) in the above index devised by Kotal et al. (2009) as shown in equation 2.

GPP = ((ξ850 x M x I)/S) * (TCHP/40)      if ξ850 >0, M >0 and I >0      Equation 2

     =   0                        if ξ850 ≤0, M ≤0 or I ≤0

Where TCHP is the tropical cyclone heat potential. Other parameters remain the same as in equation 1.

TCHP is the integrated heat content per unit area relative to the 26°C isotherm (equation 3) (Leipper and Volgenau 1972). It provides the amount of energy available in the ocean surface and subsurface for the cyclone intensification.

$$TCHP = \rho Cp \int_0^{D26} (T - 26)dz \qquad \text{Equation 3}$$

Where $C_p$ (4178 J kg$^{-1}$ °C$^{-1}$) is the heat capacity of ocean water at constant pressure, ρ (1026 kg m$^{-3}$) is the average ocean water density in the upper ocean, and Z26 is the depth of 26°C isotherm in the ocean.

## 2.2. Cooling inhibition index

Various indices are devised to measure the cyclone impact on the ocean and its energy. One such index is the cooling inhibition (CI) index defined by (Vincent et al. 2012b) which measures the amount of potential energy required to cool the ocean by 2°C through vertical mixing. It is given by equation 4.

$$CI = \sqrt[3]{[\Delta E_p(-2°C)]} \qquad\qquad \text{Equation 4}$$

Where $\Delta E_p$ is the potential energy which is a function of the mixed layer thickness and the stratification. This index is useful as it describes the oceanic inhibition to wind-induced cooling. Larger value of CI indicates a larger amount of energy is required to cool the ocean surface by 2°C.

## 2.3. Power Dissipation and accumulated cyclone energy index

Power dissipation index (PDI) and accumulated cyclone energy (ACE) are commonly used to assess the accumulated power of the cyclone. PDI is proportional to the cube of the maximum wind speed of the cyclone (Emanuel 2005). It is given by equation 5

$$PDI = \int_0^t V_{max}^3 \, dt \qquad\qquad \text{Equation 5}$$

Where $V_{max}$ is the maximum wind speed of the cyclone and t is the total duration of the cyclone.

ACE is proportional to the square of the maximum wind speed of the cyclone. It is calculated by summing the squares of the 6-hourly maximum wind speed in knots for the duration when the cyclone has a wind speed of 35 knots (64.8 km h$^{-1}$) or higher. The number is divided by 10,000 to make it more readable and easy to interpret (equation 6) (Camargo and Sobel 2005).

$$ACE = \ 10^{-4} \ \sum V_{max}^2 \hspace{3cm} \text{Equation 6}$$

## 3. Role of the ocean in modulating cyclone characteristics

3.1. Role of SSTs

3.1.1. Role of SSTs in the genesis of cyclones

Studies show that high SSTs play a major role in the genesis of cyclones in the north Indian Ocean (Kotal et al. 2009; Yokoi 2010; Chowdhury et al. 2020b). High SSTs increase the transfer of heat fluxes from the ocean to the atmosphere, which favors enhanced convection and provides conducive conditions for the genesis of a cyclone. For example, very warm SSTs (30-31°C) in the Arabian Sea prior to the genesis of Cyclone Gonu led to an enhanced heat flux transfer from the ocean to the atmosphere. This fueled the cyclone to intensify into a very severe cyclone (Krishna and Rao 2009). The genesis and rapid intensification of Cyclone Ockhi in November 2017 was also assisted by warm SSTs (>29°C) over a large area (Sanap et al. 2020; Singh et al. 2020). Similarly, the genesis of Cyclone Madi in December 2013 and that of Cyclone Phyan in November 2009 was fueled by very warm SSTs of about 29°C (Byju and Kumar 2011; Rajasree et al. 2016).

Figure 4 shows the SSTs prior to cyclogenesis in the north Indian Ocean. The SSTs were averaged for a week immediately before the day each cyclone is formed, over a 5°x5° region around the genesis center. There is a statistically significant increase in SSTs prior to cyclogenesis in both the basins, during both the cyclone seasons. The SST change is largest over the Arabian Sea, with a total change of 1.4°C in the pre-monsoon and 1.2°C in the post-monsoon season. Over the Bay of Bengal, the observed change in SST is 0.8°C in the pre-monsoon and 0.5°C in the post-monsoon.

An eastward shift in the genesis location of the very severe cyclones (wind speed ≥ 65 knots) is observed in the Bay of Bengal during the period 1997–2014 as compared to the earlier period 1981–1996, which is attributed to the rapid increase in SSTs in the south-east Bay of Bengal (Balaji et al. 2018). Further, the frequency of cyclones in the north Indian Ocean has a positive correlation with the SSTs in the eastern Indian Ocean (Yuan and Cao 2013). They show that the warming of the east Indian Ocean is associated with the negative IOD pattern. Negative IOD facilitates cyclone genesis by providing conducive ocean and atmospheric conditions for the cyclogenesis (Yuan and Cao 2013; Mahala et al. 2015). This may also be related to the fact that warm SSTs in the east Indian Ocean facilitate conducive conditions associated with the convective phase of the MJO (Singh et al. 2020).

In the Arabian Sea, cyclone genesis is generally confined to the eastern Arabian Sea (east of 70°E), due to the relatively warmer SSTs (above the threshold of 27°C) in the east Arabian Sea as compared to the west Arabian Sea. However, with the rapid warming of the west Arabian Sea, the SSTs in this region are now crossing the threshold that is required for cyclogenesis, and cyclones are now observed in this region in recent decades (Mohanty et al. 2012).

### 3.1.2. Role of SSTs in the intensification of cyclones

Recent observations indicate that cyclones in the north Indian Ocean are now exhibiting rapid intensification, intensifying by more than 50 knots in a span of just 24 hours, in response to SSTs much higher than 30°C, prominently due to the rapid warming in the region (Roxy et al. 2015, 2019). From 2000 onwards, the frequency of cyclones undergoing rapid intensification in the north Indian Ocean has increased (Vinodhkumar et al. 2021). The percentage of cyclones undergoing rapid intensification in the north Indian Ocean is higher (38%) than the cyclones in the northwest Pacific Ocean where this rate is 22% (Fudeyasu et al. 2018). Recently, in the Arabian Sea, Cyclone Vayu in June 2019 and Cyclone Nisarga in June 2020 intensified rapidly due to very warm SSTs (30-31°C) in this basin along the cyclone path. Similarly, in the Bay of Bengal, Cyclone Nargis in April 2008, intensified rapidly as it traveled over warm SSTs (>30°C) (Yu and McPhaden 2011). Cyclone Amphan in May 2020, was the first super cyclone to form in the Bay of Bengal after the 1999 Odisha super cyclone. It intensified from a severe cyclonic storm (wind speed 55 knots) to a super cyclonic storm (wind speed 120 knots) in just 24 hours. Further, it maintained the super cyclone status for nearly 24 hours. Such massive rapid intensification of Cyclone Amphan and its long duration as a super cyclonic storm was again aided by very warm SSTs (30-31°C) in the Bay of Bengal along with conducive atmospheric conditions. As discussed in previous section, due to the eastward shift of the cyclone genesis location in the Bay of Bengal, the cyclones are now traveling for a long time over the ocean and drawing more of the thermal energy released from the warm ocean waters, thereby enhancing the chances of developing into a very severe cyclone with intensity greater than 65 knots (Balaji et al. 2018). Girishkumar and Ravichandran (2012) also show that the cyclones forming east of 90°E in the Bay of Bengal generally intensify into intense cyclones as they travel over the ocean for a longer duration. In the Arabian Sea, a recent increase in the frequency of extremely severe cyclones during the post-monsoon season is linked with the warming SSTs and decreasing vertical wind shear over the region (Murakami et al. 2017). A similar response is observed over other basins such as the northwest Pacific Ocean and the southern Indian Ocean, where an increasing trend in the SSTs has assisted in the rapid intensification of cyclones in the recent decade (Mei and Xie 2016; Vidya et al. 2020).

In the pre-monsoon season, there is a high correlation ($r = 0.64$) between the SSTs and the power dissipation index that is directly proportional to the intensity of the cyclone (Sebastian and Behera 2015). They found out that there is a contrast in the correlation between the SSTs and the power dissipation index depending on the basin. The correlation is 0.73 for the Arabian Sea and 0.39 for the Bay of Bengal during the pre-monsoon season (Sebastian and Behera 2015). This shows that in the Bay of Bengal pre-monsoon season, SST alone is not a single governing factors that assists the intensification of cyclones. The reason for the difference in the correlation between the two basins is not understood yet. Regardless, it indicates that the correlation is spatially non-uniform. Hoarau et al. (2012) indicate that the correlation between the intensity of the cyclone and the SSTs is insignificant (only 0.018) for high intensity cyclones (wind speed higher than 100 knots) if we consider the entire north Indian Ocean. Based on the data for 1998–2011, Ali et al. (2013) showed that the intensity of more than 50% of the cyclones in the north Indian Ocean had no correlation with the SSTs. Similarly, a weak correlation is found between the SSTs and the cyclone intensity in the Bay of Bengal (Vissa et al. 2013a). The correlation between the SSTs and the intensity of cyclones varies within the Bay of Bengal, with a high correlation (r=0.52) in the south Bay of Bengal and a low correlation in the north Bay of Bengal (Albert and Bhaskaran 2020). This shows that the influence of SSTs on the intensity of cyclones is not uniform and that it varies spatially within the basin. These results also indicate that, other than SSTs, other factors may also be governing the intensity of cyclones in the north Indian Ocean.

3.2. Role of ocean heat content

The intensity of cyclones in the north Indian Ocean is governed not only by the SSTs but also by the high ocean heat content and warm ocean subsurface (Girishkumar and Ravichandran 2012; Patnaik et al. 2014; Maneesha et al. 2015; Sun et al. 2015). This is similar to the other ocean basins where the ocean heat content also plays a major role in governing the intensity of cyclones along with the SSTs (Wada and Usui 2007; Balaguru et al. 2013; Trenberth et al. 2018; Hallam et al. 2021).

Sharma and Ali (2014) show that ocean heat content acts as an important parameter, governing the life cycle and intensity of cyclones in the north Indian Ocean. High ocean heat content implies a warmer upper ocean, which helps cyclones to sustain or further intensify due to the uninterrupted supply of sensible and latent heat fluxes from the ocean surface to the atmosphere (Shay et al. 2000). For the period 1993–2011, Girishkumar et al. (2015) show that the accumulated TCHP in the Bay of Bengal has a positive correlation ($r = 0.62$) with the intensity of cyclones (Figure 5). The TCHP in this basin exhibits distinct interannual variability with higher TCHP during La Niña years resulting in a higher intensity and increased rate of rapid intensifications in cyclones, as compared to El Niño years (Figure 5). A composite study using the Array for Real-time Geostrophic Oceanography (ARGO) and Indian National Centre for Ocean Information Services-Global Ocean Data Assimilation System (INCOIS-GODAS) data show that in the Bay of Bengal, at least 40 kJ cm$^{-2}$ of upper ocean heat content is necessary for a cyclone to intensify (Maneesha et al. 2015; Busireddy et al. 2019). Further, in this basin, high accumulated TCHP is required for the cyclone to intensify during the pre-monsoon season, as compared to the post-monsoon season (Vissa et al. 2013a). High enthalpy fluxes in the Bay of Bengal during the post-monsoon season (~ 150 W m$^{-2}$) as compared to the pre-monsoon season (~ 100 W m$^{-2}$) favor the intensification of the cyclone even with lesser ACE (Vissa et al. 2013a). This difference in the enthalpy flux between the two cyclone seasons can be due to the difference in the mean wind speed, atmospheric humidity and barrier layer thickness in the two seasons (Vissa et al. 2013a).

Ghetiya and Nayak (2020) show that the mean TCHP for a developing cyclone (low pressure system intensifying to a cyclone) is 89.3 kJ cm$^{-2}$ in the Bay of Bengal and 89.34 kJ cm$^{-2}$ in the Arabian Sea, based on observed and reanalysis data for the period 1993–2017. On the other hand, for a non-developing low pressure system (not intensifying to a cyclone), the mean TCHP is 79.29 kJ cm$^{-2}$ in the Bay of Bengal and 75.07 kJ cm$^{-2}$ in the Arabian Sea. It is seen that a higher TCHP favors the intensification of the cyclone. However, the passage of a cyclone also reduces the TCHP by about 20-25 kJ cm$^{-2}$ in the post-monsoon season and >25 kJ cm$^{-2}$ in the pre-monsoon season within 250 km radius of the cyclone (Busireddy et al. 2019). This difference in the magnitude in the TCHP in response to the cyclone during the two seasons is linked with the different mean conditions of the ocean in the two seasons. Due to the presence of a barrier layer in the post-monsoon season in the Bay of Bengal, the reduction in TCHP due to cyclone-induced mixing is less as compared to the pre-monsoon season. The reduction in the TCHP depends on the cyclone intensity also, as intense cyclones lead to more reduction in the TCHP as compared to the weak cyclones (Busireddy et al. 2019). This shows that there is two-way feedback between the cyclone and the ocean.

Considering the importance of the TCHP in the lifecycle of the cyclone, Suneeta and Sadhuram (2018) proposed a modified genesis potential parameter (modified GPP) formula (shown in equation 2 earlier) for estimating the genesis potential of cyclones in the Bay of Bengal. By using this modified GPP, Singh et al. (2020) show that the signature of the genesis of the cyclone can be seen one day in advance as compared to the original GPP calculations (equation 1) that do not incorporate the ocean heat content. Ghetiya and Nayak (2020) show that the modified GPP is better in estimating the non – developing low-pressure systems in the north Indian Ocean.

The upper ocean content in the north Indian Ocean is in turn affected by the propagation of Kelvin and Rossby waves in this region. Genesis and intensification of Cyclone Megh in the Arabian Sea in the post-monsoon season in 2015 were associated with the downwelling Rossby waves that enhanced the upper ocean heat content and aided the formation and intensification of the cyclone (Chowdhury et al. 2020b). These waves are also affected by the ENSO conditions in the Pacific Ocean. During a La Niña, an increased upper ocean heat content in the Bay of Bengal is observed due to the downwelling coastal Kelvin wave which deepens the thermocline in the east Indian Ocean (Girishkumar and Ravichandran 2012). Thus in a La Niña year, the Bay of Bengal has higher ocean heat content as compared to the El Niño year resulting in more intense cyclones as compared to an El Niño year (Girishkumar and Ravichandran 2012).

3.3. Role of ocean eddies

The presence of mesoscale features such as eddies significantly affects the intensity of cyclones. Warm core eddies are associated with downwelling regions and have anomalously high sea surface heights and high ocean heat content (Gopalan et al. 2000; Sadhuram et al. 2004, Kumar and Chakraborty 2011). Cold core eddies are associated with upwelling regions and have anomalously low sea surface heights and low ocean heat content. In the Bay of Bengal, during the pre-monsoon season, these eddies are generated by the baroclinic instability in the coastal western boundary current (Kurien et al. 2010). These eddies are also a common feature in the post-monsoon season in the east Indian coastal current (Gopalan et al. 2000; Vinayachandran et al. 2005). The eddies in the west Arabian Sea are ~30% more in number in the pre-monsoon and post-monsoon season as compared to the southwest monsoon season (Piontkovski et al. 2019). In the north Indian Ocean as whole, warm core eddies are generally more prominent in the pre-monsoon season than the post-monsoon season (Jangir et al. 2020).

When a cyclone passes over a warm core eddy, it generally intensifies due to the high ocean heat content in the water column (Shay et al. 2000; Ali et al. 2007). On the contrary, when the cyclone passes over a cold core eddy its intensity gets reduced (Sreenivas and Gnanaseelan 2014). During the period 2001–2018, about 40% of the cyclones in the north Indian Ocean were affected by warm core eddies, resulting in an intensification of cyclones, as indicated by sea level anomaly data derived from satellite observations (Jangir et al. 2020). On the other hand, 30% of the cyclones were affected by cold core eddies, causing a weakening of cyclones (Jangir et al. 2020).

Patnaik et al. (2014) show that in the year 1999, the presence of a warm core eddy near the east coast of India intensified a post-monsoon Bay of Bengal cyclone by 260%. Such intensification was also observed for the Hurricane Opal in the Gulf of Mexico which intensified suddenly from 956 hPa to 916 hPa over 14 hours after passing over a warm core eddy (Shay et al. 2000). In the Bay of Bengal, Cyclone Aila intensified by 43% after passing over a warm core eddy, in the pre-monsoon season, as indicated by satellite and in-situ data (Sadhuram et al. 2012). However, Cyclone Jal, which was forecasted to become a very severe cyclone, suddenly weakened to a cyclonic storm category cyclone after it passed over a cold core eddy (Sreenivas and Gnanaseelan 2014). Similarly, Cyclone Madi in the Bay of Bengal in December 2013 weakened quickly after it encountered a cold core eddy (Anandh et al. 2020; Chowdhury et al. 2020a). Using numerical model experiments for five cyclones in the Bay of Bengal, Anandh et al. (2020) show that the contribution of warm core eddy in the intensification of the cyclones is more in the post-monsoon season as compared to the pre-monsoon season. Also, they note that the transfer of heat fluxes from the ocean to the atmosphere increases when a warm core eddy is present. It is hence obvious that these eddies play a crucial role in the intensification and dissipation of the cyclones in the north Indian Ocean, depending on their thermal characteristics. Similar to the north Indian Ocean, the importance of eddy in modulating the cyclone intensity is also observed in the Pacific Ocean and the Atlantic Ocean (Bao et al. 2000; Shay et al. 2000; Jullien et al. 2014).

3.4. Importance of using ocean-atmosphere coupled models for the prediction of cyclones

Earlier studies have shown that incorporating the two-way feedback between the SSTs and cyclones into the mesoscale and Weather Research and Forecast (WRF) models during the cyclone lifecycle has a significant impact on the prediction of cyclone intensity for the north Indian Ocean (Mandal et al. 2007; Bongirwar et al. 2011). Bongirwar et al. (2011) show that the improvement in cyclone intensity prediction after incorporating SSTs is due to the realistic representation of heat fluxes from the ocean to the atmosphere, which is dependent on the SSTs. Using a hurricane forecast model, Mohanty et al. (2019) show that the errors in the mean intensity reduced significantly for 12–120 hour forecast length, with realistic SSTs. They showed that the bias in the intensity forecast reduced by 50% and the landfall location of the cyclone was better forecasted when the SSTs were incorporated in the model. Similarly, based on a three-dimensional Price

Weller Pinkel model, an improvement in the intensity prediction of the cyclone by 29-47% is observed when the air-sea coupling was incorporated into the model (Srinivas et al. 2016). The improvements are mainly because of simulating the air-sea interaction processes in terms of deepening of ocean mixed layer due to cyclone induced mixing, reduction in the enthalpy fluxes, and cyclone induced SST cooling similar to the observations (Srinivas et al. 2016). For Cyclone Nilam, the improvement in the intensity of the cyclone is observed after using a three-dimensional ocean mixed layer model which improves the air-sea interaction through the realistic representation of fluxes in the model (Mohan et al. 2015). Similar to the north Indian Ocean, an improvement in the modelled intensity of cyclones is observed in the Atlantic (Sanabia et al. 2013; Kim et al. 2014; Yablonsky et al. 2015) and Pacific Ocean (Sandery et al. 2010; Jullien et al. 2014), after incorporating the coupled SSTs in the coupled model.

While the local ocean-atmosphere coupling plays an important role in cyclogenesis and its intensification, its effect on the tracks of cyclone is still unclear. In the Pacific Ocean, it is observed that air-sea coupling in an ocean-atmosphere coupled models improve the realistic representation of cyclone tracks and their spatial distribution (Jullien et al. 2014; Ogata et al. 2015), while such an effect on the tracks of cyclones is not captured by modeling studies for the Atlantic Ocean (Zhu et al. 2004; Zhu and Zhang 2006). For the north Indian Ocean, Lengaigne et al. (2018), using a regional ocean-atmosphere coupled model show that the air-sea coupling does not affect the spatial cyclogenesis pattern of the cyclone, however, it improves the bimodal seasonal distribution of the cyclone. Similarly, using the ERA-Interim, Camp et al. (2015) show that the seasonal track forecast of the north Indian Ocean cyclone is poor. Contradictory to this, model simulation in the Bay of Bengal using a three-dimensional Price-Weller-Pinkel model shows that incorporating the SST improves the track forecast of the cyclone up to 39% (Srinivas et al. 2016). Using a mesoscale

MM5 model Mandal et al. (2007) show that including realistic SST into the model improves the track prediction of the Odisha Super Cyclone which hit Odisha coast in October 1999. Similarly, using a mesoscale model, Sharma and Ali (2014) show that incorporating sea surface height anomaly in the model improves the cyclone track forecasted by the model. These studies show that air-sea coupling might play an important role in determining the track of the cyclone in the north Indian Ocean.

## 4. Ocean dynamics in response to cyclones

4.1. Ocean cooling and the role of barrier layer, mixed layer and stratification

4.1.1. Cyclone induced cooling in the Bay of Bengal: Role of ocean characteristics

In Section 3, we have seen that warm ocean surface and subsurface conditions play a dominant role in the genesis and the intensification of cyclones. The cyclone derives the energy from the ocean in terms of heat flux and moisture. On the other hand, it cools down the ocean through multiple processes. In the Bay of Bengal, the cyclone-induced cooling varies from 2℃-3℃ in the pre-monsoon season (Rao 1987; Gopalkrishna et al. 1993; Sengupta et al. 2008) to 0.5℃-1℃ in the post-monsoon season (Subrahmanyam et al. 2005; Sengupta et al. 2008). For some cyclones in the post-monsoon season, such as the Odisha super cyclone, cooling as large as 5℃-6℃ and for Cyclone Hudhud, cooling of 3℃ is observed (Sadhuram 2004; Rao 2007; Shengyan et al. 2019). Similarly, in the pre-monsoon season, Cyclone Mala in 2006 induced a large SST cooling of 4°C-5℃ (Badarinath et al. 2009; Vissa et al. 2012).

The difference in the cyclone-induced SST cooling in the pre-monsoon and the post-monsoon season is linked to the different characteristics of the north Indian Ocean in these two seasons (Figure 1). During the post-monsoon season, the Bay of Bengal has a shallow mixed layer and a thick barrier layer leading to strong stratification (Thadathil et al., 2007). However, in the pre-monsoon season, the barrier layer is thin as compared to the post-monsoon season, leading to a weaker stratification (Li 2017). The presence of such a barrier layer in the ocean inhibits the vertical mixing and suppresses the upwelling of cold subsurface waters to the surface (Sprintall and Tomczak 1992; Wang et al. 2011; Vissa et al. 2013b). Maneesha et al. (2012) show that a barrier layer formed due to the low saline water in the Bay of Bengal during Cyclone Nargis in April 2008 played a major role in the intensification of the cyclone. Using a mixed layer model, Vissa et al. (2013b) show that the presence of a thick barrier layer during Cyclone Sidr resulted in reduced SST cooling. Using observation and high resolution coupled regional climate model, Balaguru et al. (2012) show that strong salinity stratification is associated with weak cyclone-induced SST cooling and strong latent and sensible heat fluxes from the ocean to the atmosphere, which results in the increase in the intensity of the cyclones. They further suggest that, the rate of intensification of cyclones over the barrier layer region in the north Indian Ocean is 4.6 km $h^{-1}$ in 36 hours (for the period 1998–2007) which is about 50% higher than the areas where the barrier layer is absent (Balaguru et al. 2012). If we compare with other basins then it is observed that the intensification rate of cyclones over the barrier layer was highest in the southwest Pacific Ocean with an intensification rate of 9 km $h^{-1}$ in 36 hours and lowest in the northwest tropical Atlantic Ocean with an intensification rate of 3.5 km $h^{-1}$ in 36 hours (Balaguru et al. 2012). For the South Pacific Ocean, using model analysis, cyclone intensity is found to be increasing with an increase in barrier layer thickness (at a rate of 5 hPa drop in cyclone central pressure for every 10 m increase of barrier layer thickness) (Jullien et al. 2014).

Based on observational estimations for the period 1998–2004, it is found that due to the absence of a thick barrier layer during the pre-monsoon season in the Bay of Bengal, the cyclone induced cooling can reach up to 3℃ in the north Bay of Bengal (Sengupta et al. 2008). On the other hand, in the post-monsoon season, due to the presence of a thick barrier layer, the cyclone-induced cooling is only about 1℃ (Sengupta et al. 2008). A similar analysis by Neetu et al. (2012) using an ocean general circulation model also shows that there is a significant difference in the thermal stratification and upper ocean salinity of the Bay of Bengal in the post-monsoon and pre-monsoon seasons. Due to these different ocean characteristics in the two seasons, the cyclone-induced cooling is roughly three times more in the pre-monsoon season than the post-monsoon season (Figure 6). There is a difference in the estimated cyclone-induced cooling between the studies conducted by Sengupta et al. (2008) and Neetu et al. (2012). This difference is possibly due to the difference in the area considered along the track of the cyclone to estimate the cyclone induced cooling. Sengupta et al. (2008) considered 1° wide strip (~50 km radius), whereas Neetu et al. (2012) considered 4° wide strip (~200 km radius) along the track of cyclones. On average, 60% of the reduction in the cyclone-induced cooling in the post-monsoon season is contributed by thermal stratification and 40% by haline (salinity) stratification (Neetu et al. 2012). There are regional variations on how these two factors reduce cyclone induced cooling in the Bay of Bengal. In the north Bay of Bengal, it is generally the haline stratification that controls the reduction in the cyclone induced cooling, particularly due to the freshwater inflow and increased rainfall here. Whereas, in the southwest Bay of Bengal it is generally the thermal stratification that governs the reduction in cyclone-induced cooling, as these waters are characterized by the permanently warm SSTs above 28℃ (Neetu et al. 2012; Roxy et al. 2019). Due to the regional variation in the ocean subsurface characteristics (Figure 1), more cyclone-induced cooling is generally observed in the south Bay of Bengal, and the magnitude of SST cooling decreases as we head towards the north Bay of Bengal (Pothapakula et al. 2017). The absence of haline stratification in the south Bay of Bengal results in more cyclone-induced cooling in this region. A similar effect of haline stratification was observed in the Amazon-Orinoco river region in the North Atlantic Ocean where strong fresh water discharge into the ocean results in ~50% reduction in the cyclone induced cold wake as compared to the open ocean (Reul et al. 2014; Hernandez et al. 2016).

Various case studies using Biogeochemical-ARGO (BGC-ARGO) and moored buoy data during cyclones, Sidr, Thane, Vardah show that the cyclone passage also affects the mixed layer of the ocean. It is found that due to the vertical mixing induced by these cyclones, the mixed layer gets deepened by 15-30 m (Wang and Zhao 2008; Tummala et al. 2009; Chacko 2018a; Ye et al. 2018).

4.1.2. Cyclone induced cooling in the Arabian Sea: Role of ocean characteristics

Similar to the Bay of Bengal, the cyclone-induced cooling in the Arabian Sea is more in the pre-monsoon season than the post-monsoon season (Neetu et al. 2012). An average cooling of 1.2°C was observed in the pre-monsoon season and 0.6°C in the post-monsoon season, based on observed data for the period 1998–2007  (Neetu et al. 2012) (Figure 6). Multiple case studies based on observed data for cyclone 08A, Phyan, Phet and Ockhi show that cyclone induced cooling varies from 2-6°C (Premkumar et al. 2000; Subrahmanyam et al. 2005; Naik et al. 2008; Byju and Kumar 2011; Muni Krishna 2016; Singh et al. 2020). Using observations and modelling, Neetu et al. (2012) show that the cyclone induced cooling is large in the pre-monsoon season due to large cyclone wind power input. In comparison, cyclone induced cooling is weak in the post-monsoon season due to the presence of thermal stratification (Subrahmanyam et al. 2005; Neetu et al. 2012).

### 4.1.3. Cyclone induced cooling in the north Indian Ocean: Role of coupled ocean-cyclone processes

Based on the cyclone data for 1998–2007, Lloyd and Vecchi (2011) show that the cooling induced by the cyclones in the north Indian Ocean increases with an increase in the cyclone intensity only up to category 2 (wind speed up to 95 knots). From a category 1 to category 2 cyclone, the SST cooling increases from 0.6°C to 1.1°C, with a difference of 0.5°C between the categories (Figure 7). For cyclones of intensity greater than category 2, the cyclone-induced SST cooling does not increase significantly with a further increase in the intensity of cyclones. This shows that the relationship between the SSTs and the cyclone intensity is not linear. On the contrary, based on the wind power index (which is a proxy of the kinetic energy transferred by the cyclone to the ocean) Neetu et al. (2012) show that the cyclone-induced cooling keeps on increasing with the cyclone intensity. This difference is associated with the fact that the wind power index incorporates both the intensity and the translation speed of the cyclone (Neetu et al. 2012; Vincent et al. 2012b). While numerous studies have explored cyclone-induced SST cooling, only a limited have looked into the recovery process and time undergone by these SSTs to reach the pre-cyclone level. It is observed that the recovery time of the SSTs to the pre-cyclone value is not consistent and varies from six days to one month depending on the magnitude of the SST cooling (Mandal et al. 2018).

Cyclone rainfall also has a significant effect on the cyclone induced cooling (Jacob and Koblinsky 2007; Jourdain et al. 2013). Jourdain et al. (2013) using data for the period 2002–2009, show that the cyclone rainfall tends to reduce the cyclone induced cold wake due to the stabilizing effect of rainfall that reduces the vertical mixing in the ocean induced by the cyclone. This reduction in the cold wake due to cyclone rainfall is maximum in the Arabian Sea among all the basins due to the strong thermal stratification which reduces the vertical mixing, leading to a reduction of entrainment of cooler water from the subsurface to the surface (Jourdain et al. 2013). The Bay of Bengal has a smaller effect in response to the cyclone rainfall on the reduction in the cyclone induced cold wake because of the presence of a strong barrier layer, as discussed earlier.

The cooling induced by cyclones in the north Indian Ocean is mainly governed by the vertical mixing induced by the cyclone winds (Murty et al. 1983; Girishkumar et al. 2014; Prakash and Pant 2017; Kumar et al. 2019). This is in line with other basins as well, where also cyclone-induced cooling is mainly governed by the vertical mixing and cyclone-induced upwelling, as compared to other processes such as the air-sea flux exchange and advection (Price 1981; Jacob et al. 2000; Huang et al. 2009; Jullien et al. 2012). Observations and coupled ocean-atmosphere model experiments show that during Cyclone Jal (November 2010), that formed in the Bay of Bengal, the heat fluxes and horizontal advection together contributed alone about 33% in the cyclone induced cooling. The remaining contribution in the cooling was mainly from vertical mixing (Girishkumar et al. 2014). Similarly, Kumar et al. (2019) using observations and model experiments show that during Cyclone Roanu, the cooling due to air-sea fluxes contributed to only about 25% of the total cyclone-induced cooling, highlighting that cyclone induced vertical mixing plays a dominant role in the cooling. This is similar to the findings of Vincent et al. (2012a) in which they used model experiments for the period 1978–2007 to show that for strong cyclones, vertical mixing and ocean mean state plays dominant roles in cyclone-induced cooling. On the contrary, for Cyclone Sidr in the north Indian Ocean, Vissa et al. (2013b) using a mixed layer model show that when a barrier layer is present, then the SST cooling is mainly driven by air-sea heat fluxes. On the other hand, in the absence of the barrier layer, the SST cooling is mainly driven by the vertical mixing and entrainment flux. Similarly for the period 1978–2007, using an ocean general circulation model, Neetu et al. (2012) show that vertical mixing plays a major role in SST cooling in the pre-monsoon when the barrier layer is absent. However, in the post-monsoon season, when the barrier layer is present, latent heat fluxes play a major role in the cyclone induced SST cooling (Neetu et al. 2012). The factors controlling the cyclone induced SST cooling also varies during the lifecycle of the cyclone. Based on observational analysis for Cyclone Chapala, Chowdhury et al. (2020b) show that during the initial phase of the cyclone, the SST cooling is mainly due to the enhanced evaporation due to the strong cyclone winds. However, in the decay phase, along with the evaporative cooling, the vertical mixing and cyclone-induced Ekman pumping also contributes to the enhanced cooling after the passage of the cyclone (Chowdhury et al. 2020a).

Interestingly, there are cyclone-ocean interactions in the north Indian Ocean that do not cool the ocean surface. During post-monsoon season in the Bay of Bengal, Cyclone Thane (2011) has increased the SSTs by 0.6°C (Mathew et al. 2018). Using moored buoy observations, Chacko (2018) shows that this warming after the passage of cyclone occurred in the upper 40 m of the ocean, whereas a cooling was observed in the depths of 40-80 m. This contrasting response of the ocean is because of the presence of an inversion layer prior to the cyclone. As the cyclone moves over the area of the inversion layer, warmer ocean subsurface water comes to the surface due to cyclone-induced upwelling, leading to an increase in the SSTs after its passage.

4.1.4. Cyclone induced cooling in the north Indian Ocean: Role of cyclone translation speed

Cyclone induced cooling in the north Indian Ocean is also dependent on the translation speed of cyclones. The slow-moving cyclones (average speed of 9 km h$^{-1}$, 25$^{th}$ percentile) cause relatively stronger cooling as compared to the fast-moving cyclones (average speed of 16.7 km h$^{-1}$, 75$^{th}$ percentile) (Rao 2007; Mandal et al. 2018; Busireddy et al. 2019; Nigam et al. 2020). This is similar to other tropical basins where the cooling induced by the cyclone decreases with the increase in the translation speed (Dare and Mcbride 2011). In the north Indian Ocean, intense cyclones move faster as compared to the less intense cyclones (Mohapatra and Sharma 2019) which is also similar to other basins (Mei et al. 2012). In the case of Cyclone Nilofar in the Arabian Sea, stronger cooling is observed when the cyclone has a lower translation speed (Ali et al. 2020). This is because slow-moving cyclones have more time to interact with the ocean surface, leading to an increase in the upwelling and vertical mixing for a longer time that mixes the upper-ocean water column. Maneesha et al. (2019) used a case study to show that a fast-moving cyclone can overcome the negative effect of a cold core eddy and the cyclone can maintain its intensity despite passing over a cold core eddy. In the Bay of Bengal, the cyclone induced cooling in the radius 50-100 km from the cyclone center varies with the translation speed of the cyclone, as shown by a composite analysis for the period 2006–2013 (Pothapakula et al. 2017). For pre-monsoon cyclones, a cooling of 2°C for slow-moving cyclones (translation speed less than 14 km h$^{-1}$), and a cooling of 1.2°C for moderate moving cyclones (translation speed between 15-24 km h$^{-1}$), and cooling of 0.9°C for fast-moving cyclones (translation speed more than 24 km h$^{-1}$) is observed within the radius of 50-100 km of cyclone center (Pothapakula et al. 2017).

4.2. Effect of cyclone translation speed on its intensification

Further, the translation speed of cyclones plays a vital role in determining how much ocean heat content will be necessary for the cyclone intensification (Lin et al. 2009). They show that slow-moving cyclones require more upper ocean heat content to intensify to a category 5 cyclone as compared to the fast-moving cyclones. Similar to the northwest Pacific, a slow-moving cyclone in the Bay of Bengal also requires higher ocean heat content for intensification than the faster moving cyclones (Sadhuram et al. 2010). This is because a slow-moving cyclone gets affected by the ocean negative feedback for a longer time leading to a decrease in the intensity of the cyclone. Therefore, for a slow-moving cyclone, a large amount of ocean heat content is required for cyclone intensification. Meanwhile, Sadhuram et al. (2010) show that fast-moving cyclones can intensify even with low ocean heat content as the cyclone will not get affected by the ocean negative feedback as the cyclone moves away before the ocean heat content decreases. The magnitude of the translation speed required for intensification depends on the ocean heat content and is given by the empirical formula shown below (equation 7) (Sadhuram et al. 2010).

$(U_h = -0.03*UOHC + 6.1)$                     Equation 7

Where $U_h$ is the translation speed of the cyclone and UOHC is the upper ocean heat content. Sadhuram et al. (2010) tested this formula with Cyclone Sidr and Nargis and the computed results of the translation speed were in good agreement with the observed values. The studies here point out that the intensification of cyclones is affected by the translation speed of cyclones.

4.3. Asymmetric response in ocean cooling and currents to the cyclone

The cooling induced by the cyclone along the track of the cyclone is not symmetrical (Rao 2007; Ye et al. 2018, Wang et al. 2012b). More cooling is induced on the right-hand side of the cyclone track in the open ocean (McPhaden et al. 2009; Wang and Han 2014). This is because, in an open ocean, as the cyclone moves forward, the associated wind stress veers in a cyclonic direction at any given point along the right-hand side of the cyclone track. There is an asymmetry in the wind stress due to the combination of the cyclonic winds and the translation speed with higher wind stress on the right-hand side of the cyclone center. In the Northern Hemisphere, the Coriolis force also acts on the ocean currents in the same direction of the wind stress, on the right-hand side of the cyclone track. This stronger wind stress and exchange of energy from the cyclone and to the ocean currents leads to an enhancement of the inertial currents on the right-hand side of the cyclone track, amplifying entrainment and more SST cooling on the right-hand side of the cyclone track (Price 1981).

The inertial oscillation induced by cyclonic winds have a periodicity of $2\pi/f$ which drives the near surface water in a clockwise direction, governed by the Coriolis force and gravity force (Price et al. 1994). The amplitude of these inertial currents amplitude varies depending on the strength of the cyclone winds and their periodicity, and the diameter varies with the latitude (with higher values equatorward) (Venkatesan et al. 2014). From various case studies on cyclones in the north Indian Ocean, it is observed that these inertial oscillations have a periodicity of 1.5-2.1 days (Joseph et al. 2007; Venkatesan et al. 2014; Mandal et al. 2018). These inertial currents are initially triggered at the ocean surface due to the wind forcing and later they propagate downward into the ocean subsurface (Leaman and Sanford 1975; Kundu 1976; Jarosz et al. 2007) and also move outwards, away from the cyclone wind forcing (D'Asaro et al. 2007; Mukherjee et al. 2013).

The vertical profile of the temperature of the ocean up to 100 m is significantly altered by these inertial oscillations (Rao et al. 2010). During a cyclone in 1995, inertial currents are observed as far as 300 km from the cyclone center, attaining a maximum speed of 0.75 m s$^{-1}$ (Saji et al.

2000). For very intense cyclones such as Cyclone Phailin, very intense currents of velocity 1.29 m s[-1] are observed (Venkatesan et al. 2014). During Cyclone Nargis, enhanced cooling is observed on the right-hand side of the cyclone track in the open ocean (McPhaden et al. 2009). This right-hand side enhanced cooling is due to strong inertial currents of magnitude 0.9 m s[-1] as measured by the Research Moored Array for African-Asian-Australian Monsoon Analysis and Prediction (RAMA) buoy on the right-hand side of the cyclone track (Maneesha et al. 2012). A similar enhanced right-hand side cooling of the track of the cyclone for the post-monsoon Bay of Bengal cyclones (1999) is observed by Wang et al. (2012a), using the Hybrid Coordinate Ocean Model (HYCOM). Using a simple ocean model for a Bay of Bengal cyclone, Behera et al. (1998) show that the vertical shear associated with the inertial currents in the mixed layer plays a decisive role in the surface cooling and deepening of the mixed layer. The western boundary current along the west coast of the Bay of Bengal got enhanced by ~0.5 m s[-1] due to the effect of Cyclone Roanu (Mandal et al. 2018). Similarly, in the Arabian Sea, during Cyclone Phyan, enhanced open ocean cooling was observed on the right-hand side of the cyclone track (Byju and Kumar 2011). There also inertial and sub-inertial oscillations are triggered by the cyclone winds which persists for 2-14 days (Price 1981; Rao et al. 2010; Wang et al. 2012b; Girishkumar et al. 2014).

Various observational studies show that the asymmetry in cyclone induced cold wake in the north Indian Ocean is also observed in the Atlantic and the northwest Pacific Ocean (Shay and Elsberry 1987; Shay et al. 1989, 1992; Monaldo et al. 1997; Liu et al. 2007). Using a two-layer reduced gravity ocean model for the cyclone in the Atlantic basin, Samson et al. (2009) show that the asymmetric cyclone-induced cold wake is also dependent on the translation speed of cyclones. Fast-moving cyclones lead to higher asymmetric cold wake as compared to slow-moving cyclones.

Similar results are shown by Mei and Pasquero (2013) based on a composite study using satellite data. However, an independent analysis for the north Indian Ocean is pending.

Contradictory to the open ocean rightward enhancement in the ocean cooling, near the east coast of India in the Bay of Bengal, more cooling is observed on the left-hand side of the cyclone track. This contrasting response in cyclone-induced SST cooling along the western Bay of Bengal as compared to the open ocean is because of the off-shore transport of the ocean water, leading to an enhanced upwelling towards the left of the cyclone center (Mahapatra et al. 2007). A similar enhanced cooling on the left-hand side of the cyclone center near the coast was observed in the case of Cyclone Roanu in May 2016 (Mandal et al. 2018) and Cyclone Thane in December 2011 (Chacko 2018a).

### 4.4. Cyclone induced ocean heat transport

The globally averaged cyclone-induced ocean heat uptake is found to be 0.48 PW per year, based on SST anomalies derived from the Tropical Rainfall Measuring Mission's (TRMM) Microwave Imager (TMI) for the period 1998–2005 (Sriver et al. 2008). A separate analysis for the South Pacific Ocean shows that the global average estimates of cyclone-induced ocean-heat uptake might be an overestimation and for the South Pacific Ocean it is only 0.015 PW per year (Jullien et al. 2012). Emanuel (2001) suggests that cyclone-induced mixing pumps ocean surface heat downward into the thermocline due to winds induced turbulent mixing, and may be responsible for a significant amount of oceanic meridional heat transport. However, not all of the heat pumped by the cyclone into the ocean is transported poleward, some is lost into the atmosphere (Pasquero and Emanuel 2008). Also, the poleward transportation of cyclone pumped heat may be significantly reduced due to the seasonal deepening of the mixed layer depth (Jansen et al. 2010). Using observational analysis, it is estimated that about 15% of the ocean heat transport from the tropics to the extratropics is associated with the cyclone induced downward ocean heat pumping which is then transported poleward (Sriver and Huber 2007).

Wang et al. (2012a), using a hybrid coordinate ocean model, show that two Bay of Bengal cyclones (1999 post-monsoon season) significantly reduced the ocean heat content (maximum reduction of 160 x $10^{18}$ J), due to a southward ocean heat transport out of the Bay of Bengal and reduced downward net heat flux anomalies. The southward transport of the ocean heat out of the Bay of Bengal is transported along the western boundary of the basin, and continues up to months even after the dissipation of the cyclone (Wang et al. 2012a). This shows that cyclones in the north Indian Ocean can significantly induce poleward heat transport out of the basin.

4.5. Cyclone induced changes in ocean salinity

The sea surface salinity is also affected significantly by cyclones in the Bay of Bengal and the Arabian Sea, with the salinity generally increasing after the passage of the cyclone (Maneesha et al. 2012; Mandal et al. 2018; Chaudhuri et al. 2019; Venkatesan et al. 2020). After the cyclone passes, the Bay of Bengal has a saltier wake as compared to the Arabian Sea (Jourdain et al. 2013), due to the strong haline stratification in the Bay of Bengal (Neetu et al. 2012). In the Arabian Sea, the weak response of the salinity to the cyclone passage is similar to the northeast Pacific Ocean where also the absence of haline stratification leads to a less salty wake after the cyclone passage (Jourdain et al. 2013). During Cyclone Phailin (post-monsoon season), the sea surface salinity increased by as large as 3-5 psu due to the cyclone induced vertical mixing (Qiu et al. 2019). This cyclone induced vertical mixing caused the saline subsurface water to advect to the surface, causing an increase in the surface salinity (Chaudhuri et al. 2019). During the passage of Cyclone Jal and Cyclone Roanu, an increase in salinity of 1 psu is observed (Girishkumar et al. 2014; Mandal et al. 2018). However in the case of Cyclone Amphan (May 2020), a decrease in salinity is observed near the coast of Bangladesh during cyclone passage (Chauhan et al. 2021). This decrease in salinity during cyclone passage is because of enhanced freshwater runoff due to the very heavy rainfall caused by the cyclone (Chauhan et al. 2021). A similar upper-ocean freshening due to the very heavy rain associated with the cyclones is observed in other ocean basins as well (Liu et al. 2007, 2020; Bond et al. 2011; Lin and Oey 2016; Hsu and Ho 2019). Chacko (2018b) observed that similar to the asymmetric response of the cyclone induced cooling, the salinity response to the cyclone is also asymmetrical. Their study on Cyclone Vardah using microwave surface measurements of salinity in the Bay of Bengal showed a large increase in the salinity on the right of the cyclone track as compared to the left-hand side. They attributed this asymmetry in salinity to the asymmetric rainfall caused by the cyclone, with more rainfall on the left-hand side of the cyclone center. In addition, increased vertical mixing on the right-hand side of the cyclone due to stronger winds results in significant increase in salinity. The asymmetry in cyclone related rainfall and the vertical mixing leads to less saltier wake on the left-hand side of the cyclone track as compared to the right-hand side of the cyclone track (Chacko 2018b). This asymmetry in the rainfall is observed for other north Indian Ocean cyclones also (Raghavan and Varadarajan 1981; Thakur et al. 2018; Uddin et al. 2021). Based on a statistical assessment using satellite data, it is observed that the asymmetry in rainfall arises due the atmospheric wind shear with intense cyclonic rainfall concentrated in the downshear left quadrant of the cyclone (Uddin et al.

2021)right-hand side. Such asymmetry in the cyclone rainfall is also observed in cyclones in the Northwest Pacific and the Atlantic Ocean (Cecil 2007; Xu et al. 2014; Yu et al. 2015).

4.6. Negative feedback of cyclone-induced cooling on the cyclones

The cooling induced by the cyclone has a significant negative feedback on the cyclone intensity (Bender and Ginis 2000; Zhu and Zhang 2006; Walker et al. 2014; Scoccimarro et al. 2017; Chen et al. 2018), mainly due to the reduction in the surface enthalpy fluxes (Bender et al. 1993; Schade and Emanuel 1999; Cione and Uhlhorn 2003). The negative feedback due to the cyclone-induced cooling is stronger in regions where the mixed layer is shallow or in the absence of barrier layer (Jullien et al. 2014). In the north Indian Ocean, using an ocean-atmosphere coupled and de-coupled experiment in regional model, Neetu et al. (2019) show that air-sea coupling reduces the pre-monsoon cyclone intensity. This is because intense cyclones induce large SST cooling in the pre-monsoon season, which reduces the enthalpy fluxes from the ocean to the atmosphere. However, this negative effect of air-sea coupling on cyclone intensity is offset by the stronger positive effect of seasonal changes in large-scale environmental parameters such as wind shear, and mid-level humidity. Further, they show that the effect of air-sea coupling on the contrast between the intensity of the pre-monsoon and post-monsoon cyclones is smaller as compared to the contrast in the environmental variables in these two seasons (Neetu et al. 2019). Similarly,  using a three-dimensional coupled model, Srinivas et al. (2016) show that, the coupling with the ocean significantly reduces the cyclone intensity in the Bay of Bengal by reducing the enthalpy fluxes up to 800 W m$^{-2}$. Observations show a rapid weakening of Cyclone Ockhi immediately (1 day) after the cyclone interaction with its induced cold wake, resulting in a wind speed reduction from 85 knots to 30 knots in a span of 48 hrs. (Singh et al. 2020).

## 5. Ocean biology in response to the cyclone

Cyclones have a significant influence on phytoplankton blooms, because the high cyclone winds generate strong upwelling, which enhances the phytoplankton blooms through upward Ekman pumping of nutrients into the euphotic zone (Zhang 1993; Babin et al. 2004; Prasad and Hogan 2007; Zhang et al. 2018). The biological response of the ocean to the cyclones depends on multiple factors such as cyclone intensity, its translation speed and the ocean subsurface characteristics (Price 1981; Zheng et al. 2008; Chacko 2018a).

The Bay of Bengal is generally a low productive basin as compared to the Arabian Sea due to large fresh water discharge in the Bay of Bengal leading to high stratification that suppresses the productivity in this basin (Vinayachandran et al. 2002). In the Arabian Sea, strong winds in the southwest monsoon season enhance the productivity in the west Arabian Sea (Ryther et al. 1966) though the east Arabian Sea is less productive. In the winter season, the dry northerlies cause convective mixing in the north Arabian Sea which enhances the productivity in this region (Madhupratap et al. 1996). In the pre-monsoon season (April-May), the Arabian Sea is generally nutrient-depleted with surface chlorophyll concentration varying from 0.1 to 0.4 mg m$^{-3}$ (Prasanna Kumar et al. 2000). During this time, a maximum chlorophyll concentration is observed in the subsurface at about 40–60 m depth (Bhattathiri et al. 1996).

The average surface chlorophyll concentration associated with cyclone induced phytoplankton blooms based on satellite estimates during the period 1997–2019 is about 1.65 mg $m^{-3}$ in the north Indian Ocean (Kuttippurath et al. 2021). Figure 8 shows the chlorophyll bloom induced by different cyclones in the Bay of Bengal. Based on various case studies, it is observed that the passage of a cyclone significantly increases the surface chlorophyll by 3–8 mg $m^{-3}$ (Subrahmanyam et al. 2002; Naik et al. 2008; Ali et al. 2020) and the increase in chlorophyll is ten to twenty-eight times more than the pre-cyclone chlorophyll values (Wang and Zhao 2008; Ali et al. 2020). This increase in the chlorophyll is due to the strong vertical mixing and upwelling induced by the cyclone. The strong mixing and upwelling eliminate the sub-surface maxima of the chlorophyll and it gets distributed uniformly throughout the mixed layer (Subrahmanyam et al. 2002).

In the Bay of Bengal large chlorophyll blooms ranging from 4-10 mg $m^{-3}$ are observed after the cyclone passage (Reddy et al. 2008; Sarangi 2011; Chacko 2017). Vinayachandran and Mathew (2003) show that more chlorophyll is observed for years when there is a cyclone in the post-monsoon season in the Bay of Bengal, as compared to years when there is no cyclone. This shows that cyclone is one of the major drivers that enhances the annual mean surface chlorophyll in the Bay of Bengal.

Menkes et al. (2016) show that the cyclones affect the annual primary productivity mainly in the west Bay of Bengal and the east Arabian Sea, using observed data for the period 1998–2007. The annual contribution of the cyclone in increasing the primary productivity in these areas varies between 5-10%. The chlorophyll bloom induced by the cyclone throughout the ocean is not uniform, it varies from coast to open ocean with more blooms generally seen nearer to the coast (Wang and Zhao 2008; Sarangi et al. 2014; Menkes et al. 2016). Also, few case studies in the Arabian Sea and the Bay of Bengal show that the presence of cyclonic eddies along the path of the cyclone affects the chlorophyll blooms, with the cyclonic eddies enhancing the blooms (Chacko 2017; Lü et al. 2020; Vidya et al. 2021). The presence of a cyclonic eddy in the track of the cyclone enhances the cyclone-induced upwelling, resulting in intense chlorophyll bloom after cyclone passage (Ye et al. 2018). Based on a case study for Cyclone Phyan, Byju and Prasanna Kumar (2011) show that the biological response of the ocean to the cyclone lags by 3-4 days as compared to the physical response. This is because for Cyclone Phyan, the cyclone induced Ekman pumping velocity was 2 x $10^{-4}$ m $s^{-1}$, which means that it takes about 3.4 days for the nutrients to upwell from a depth of 60 m to the surface. Once the cyclone passes away, then the increase in sunlight and availability of nutrients enhances the biological productivity. As a result, the increase in chlorophyll was seen at a lag of 3-4 days. On an average, a similar lag in the chlorophyll response due to cyclones is observed for all the tropical cyclones during the period 1998–2007 (Menkes et al. 2016).

## 5.1. Role of ocean stratification on cyclone-induced chlorophyll blooms

It is seen that during the post-monsoon season in the Bay of Bengal, stratification also plays a dominant role in controlling the cyclone induced chlorophyll blooms (Vidya et al. 2017). Depending on the strength of the cyclone and the magnitude of stratification, the cyclone can break the stratification and bring in nutrients rich water onto the surface (Vidya et al. 2017; Maneesha et al. 2019). Weak stratification along the track of Cyclone Thane enhanced the chlorophyll bloom despite its low intensity. On the other hand, strong stratification before Cyclone Phailin limited the chlorophyll bloom after the passage of the cyclone despite of its high intensity (Vidya et al. 2017; Navaneeth et al. 2019). Strong stratification also limits the spatial extent of the chlorophyll bloom in the Bay of Bengal (Patra et al. 2007). Maneesha et al. (2019) show that there is an inverse relationship between the barrier layer and cyclone-induced chlorophyll blooms. A shallow mixed layer also facilitates the enhancement of chlorophyll blooms during cyclone passage (Chacko 2019). Chacko (2019) compared the ocean subsurface conditions during cyclones Hudhud and Vardah in the Bay of Bengal and concluded that the shallow mixed layer before Cyclone Hudhud played an important role in enhancing the chlorophyll blooms during the cyclone. On the other hand, a deeper mixed layer before Cyclone Vardah suppressed the chlorophyll blooms (Chacko 2019). However, in the coastal areas once the cyclone has made landfall, the inverse relationship between the barrier layer and cyclone-induced chlorophyll blooms does not hold. In coastal areas, after the cyclone landfall, productivity tends to rise from land-driven nutrients due to the large amount of freshwater influx triggered by coastal flooding (Maneesha et al. 2019).

5.2. Role of cyclone intensity and translation speed on the cyclone-induced chlorophyll blooms

Based on a case study for Cyclone Phailin, Latha et al. (2015) show that there is a positive correlation between the intensity of the cyclone and the cyclone induced chlorophyll bloom. However, based on long term data for the period 1990–2016, Chacko (2019) shows that the chlorophyll bloom is not only affected by the intensity of the cyclone in the north Indian Ocean. They show that there is an inverse relationship between the translation speed of the cyclone and the increase in the chlorophyll concentrations, with a significant correlation ($r = -0.69$). A similar study in the north Indian Ocean using the ocean colour measurements from satellite (1997–2019) confirms the inverse relationship of the translation speed of the cyclone and the chlorophyll blooms (Kuttippurath et al. 2021). This indicates that with the decrease in the translation speed of the cyclone, the chlorophyll blooms increase. Cyclone Hudhud and Mala had similar wind speed, but Cyclone Hudhud had a lower translation speed than Cyclone Mala. As a result, Cyclone Hudhud induced stronger chlorophyll blooms than Cyclone Mala (Chacko 2019). Similarly, Cyclone Thane induced high chlorophyll in the central Bay of Bengal when the cyclone was moving with a lower translation speed (Chacko 2018a). This relationship of enhanced chlorophyll bloom with the cyclone translation speed is because slow-moving cyclones induce more momentum into the ocean surface and subsurface, and mix the ocean for a longer time as compared to the fast-moving cyclones (Sun et al. 2010; Mei et al. 2015). These studies indicate that not all intense cyclones induce chlorophyll bloom in the north Indian Ocean and even weak cyclones can induce strong chlorophyll bloom depending on the ocean subsurface characteristics and the cyclone translation speed.

## 6. Coupled interactions between cyclones and the climate system

6.1. Ocean-cyclone-atmosphere coupled interactions in the north Indian Ocean

The cyclones in the north Indian Ocean have a significant impact on the climate system also. Sreenivas et al. (2012) show that Bay of Bengal cyclones during the post-monsoon (1993–2009) significantly affects the intensity and phase propagation of Wyrkti jets. Wyrkti Jets are strong narrow equatorial eastward surface currents during the inter-monsoon periods (Wyrtki 1973) that play an important role in the zonal redistribution of mass on the seasonal time scale (Reppin et al. 1999) and affects the upper ocean heat content that alters the air-sea interaction in the Indian Ocean (Behera et al. 2000). The surface wind speed gets enhanced by 3 m s$^{-1}$ due to the cyclone which leads to an average monthly increase of the Wrykti jets by 0.4 m s$^{-1}$. Also, the post-monsoon cyclones in the Bay of Bengal increase the upper ocean heat content in the eastern equatorial Indian Ocean by enhancing the subsurface ocean currents and lead to a deepening of the thermocline.

This in turn leads to an increase in the transport of warm water to the east Indian Ocean, resulting in a subsurface warming of about 1.4ºC there (Sreenivas et al. 2012).

The severe cyclones during the pre-monsoon season in the Bay of Bengal are also found to play a role in triggering a positive IOD event in the Indian Ocean (Francis et al. 2007). The positive IOD events during the period 1982–2003 are found to be preceded by at least one severe cyclone in the Bay of Bengal (Francis et al. 2007). The severe cyclones in the Bay of Bengal in April-May strengthen the meridional pressure gradient over the eastern equatorial Indian Ocean, leading to an anomalous southeasterly flow along the Sumatra coast and enhanced coastal upwelling, causing a decrease in SSTs in this region. This leads to the development of the east-west gradient of SSTs. Further, the severe cyclones in April-May consume a large amount of atmospheric moisture leading to the anomalous lower integrated water vapor after a cyclone which causes suppression of the convection over the eastern equatorial Ocean. Francis et al. (2007) argue that this suppression of convection in the eastern equatorial Indian Ocean leads to the enhancement of the convection over the western equatorial Indian Ocean. The shift in the convection pattern leads to the weakening of the westerlies in the central equatorial Indian Ocean. This triggers positive feedback with more convection in the western equatorial Indian Ocean, suppressed convection in the eastern equatorial Indian Ocean, and anomalous easterly flow along the central equatorial Indian Ocean (Francis et al. 2007), leading to an IOD.

6.2. Impact of north Indian Ocean cyclones on the atmospheric circulation in the South China Sea

In 2019, Cyclone Fani which formed in the Bay of Bengal in late April caused an early advancement of the monsoon over the South China Sea (Liu and Zhu 2020). Cyclone Fani released a tremendous amount of latent heat which shifted the South Asian High northwards from its normal position and simultaneously strengthened the upper-level trough west of the Tibetan Plateau. Also, Cyclone Fani transported a large amount of moisture to the South China Sea which strengthened the mid-latitude trough. A combination of these factors led to an early onset of monsoon over the South China Sea in 2019 (Liu and Zhu 2020).

## 7. Observed and projected changes in the ocean-cyclone interactions

7.1. Observed changes in ocean-cyclone interactions

7.1.1. Changes in the ocean-cyclone interactions in the Arabian Sea

Observations show that the average ACE of the pre-monsoon cyclones in the Arabian Sea almost doubled from 3.8 knots$^2$ in the period 1980–1999 to 7.19 knots$^2$ in the period 2000–2019 (Figure 9a). Similarly, in the post-monsoon season, the average ACE of the cyclones has also doubled, from 2.55 knots$^2$ in the period 1980–1999 to 5.44 knots$^2$ in the period 2000–2019 (Figure 9a). . Similar to the changes in ACE, the average lifetime maximum intensity of the cyclones in the Arabian Sea has increased from 54.2 knots to 75 knots for the pre-monsoon season and from 53.2 knots to 65 knots for the post-monsoon season, between the earlier (1980–1999) and recent (2000–2019) periods (Figure 9b). This is in line with previous studies which show that the intensity of cyclones in the Arabian Sea has increased in the recent decade (Evan et al. 2011; Murakami et al. 2017; Deshpande et al. 2021).

The increasing intensity of the Arabian Sea cyclones in the recent decade is linked to an increase in SSTs and a decrease in the vertical wind shear (Evan et al. 2011; Murakami et al. 2017). However, for a long-term period involving the pre-satellite period, it is observed that the intensity of the cyclones in the Arabian Sea exhibits a multidecadal variability (Rajeevan et al. 2013). During the period 1955–1973, six intense cyclones were observed, the total number of intense cyclones decreased to four during the period 1974–1992 and again the number increases to seven during the period 1993–2011 (Rajeevan et al. 2013). This variability in the intensity of the Arabian Sea cyclones is linked to the multidecadal variability of upper ocean heat content. The upper-ocean heat content was large during the epoch 1955–1973 and 1993–2011 leading to intense cyclones in these epochs (Rajeevan et al. 2013).

### 7.1.2. Changes in the ocean-cyclone interactions in the Bay of Bengal

In the earlier section, it is seen that in the Arabian Sea, the ACE has increased significantly in the recent period (2000–2019), however in the Bay of Bengal the changes in the ACE are negligible (Figure 9a). Similarly, the cyclone maximum intensity in the Bay of Bengal (in both the pre-monsoon and post-monsoon seasons) remains almost unchanged, with a slight decrease in the lifetime maximum intensity of the cyclone is observed in the pre-monsoon season in the recent period (Figure 9b). There is no significant change in the lifetime maximum intensity of Bay of Bengal cyclones, however, it is observed that in the recent decade (1998–2015), there is a fourfold increase in the rapid intensification rate of cyclones in the post-monsoon season (Fan et al. 2020). This increase in the rapid intensification is associated with the northeastward shift of the cyclone tracks. Due to this northeastward shift in the cyclone track, now more cyclones are reaching the northern Bay of Bengal that has low salinity and is highly stratified. Due to the higher stratification, the cyclone-induced cold wake gets suppressed, resulting in favorable ocean conditions for the cyclone to gain more energy and undergo rapid intensification (Fan et al. 2020). Also, the conversion rate of cyclones intensifying into a category 3 or higher intensity cyclone east of 90°E has increased from 14% during 1981–1995 to 42% during 1996–2010 due to rapid warming of SSTs in the east Bay of Bengal (Balaguru et al. 2014).

The warming in the Bay of Bengal is possibly associated with the changes in the ocean heat content which is governed by the heat entering from the eastern side through the Indonesia throughflow (Sprintall and Revelard 2014) and the heat leaving from the western side. During the period 1999–2014, the incoming heat flux from the east into the Bay of Bengal increased significantly (from $0.18 \times 10^{12}$ W/month to $0.57 \times 10^{12}$ W/month), indicating that more heat is entering the region (Anandh et al. 2018). On the other hand, a significant reduction in the outgoing heat flux along the western boundary has also been found from the analysis. This results in the accumulation of heat in the Bay of Bengal leading to an increase in the ocean heat content (Anandh et al. 2018).

The monsoon season (June-September) generally does not see strong cyclones in the north Indian Ocean due to strong shear. However, it is observed that there is an increasing trend in the intensity and frequency of the cyclones in the monsoon season based on the satellite data until the year 2004 (Rao et al. 2008; Krishna 2009). This increase in the intensity of cyclones is linked to the warming of the tropical Indian Ocean which plays a role in the reduction of the atmospheric wind shear leading to an increase in the intensity of the cyclones in the monsoon season in recent years (Krishna 2009).

## 7.2. Projected future changes in the ocean-cyclone interactions

Few studies have examined the future changes in cyclone activity in the north Indian Ocean using the Coupled Model Intercomparison Project Phase 5 (CMIP5) model experiments (Bhatia et al.

2018; Bell et al. 2020). CMIP5 involve more than fifty models, that provide century-scale future simulations of ocean and atmosphere for different scenarios of global warming and radiative forcing. These long-term experiments help us to understand how the ocean and atmosphere will change with the increase in emissions and the radiative forcing (Taylor et al. 2012). Bhatia et al. (2018) using model simulations (RCP 4.5 scenario) projects that there will be a fourfold increase in the percentage of cyclones intensifying to an intense cyclone (category 4 or higher) in the north Indian Ocean during the period 2081–2100 as compared to the period 1986–2005. Similarly, in a 2°C warming scenario, a 5% increase in the intensity of cyclones in the north Indian Ocean is projected (Knutson et al. 2020). Not only the intensity of cyclones but also the frequency of cyclones is projected to increase by 40~50% especially in the Arabian Sea. On the contrary in the Bay of Bengal, the frequency of cyclones is projected to decline by ~30% in a global warming scenario (Murakami et al. 2013; Bell et al. 2020). With the projected decrease in the frequency of the Bay of Bengal cyclones, the landfall rate of the cyclones at the east coast of India is also projected to decrease by two cyclones per decade (Bell et al. 2020). The north Indian Ocean is rapidly warming (Roxy et al. 2019) and has contributed to more than 25% of the total increase in the ocean heat content globally in the last 20 years (Lee et al. 2015; Cheng et al. 2017). In a global warming scenario, an increase in SSTs at a faster rate in the Arabian Sea as compared to the Bay of Bengal is one of the major thermodynamic parameters due to which models are projecting an increase in the frequency of the cyclones in the Arabian Sea (Murakami et al. 2012, 2013).

Based on the TRMM estimates for the cyclones during the period 1997–2017, Ankur et al. (2020) show that the rainfall rate in the inner core (0-100 km) of the cyclone is high in the Bay of Bengal basin (9-10 mm hr[-1]) as compared to the Arabian Sea basin (7-8 mm hr[-1]). Also, the rainfall rate in the inner core of the cyclones increases with the intensity of the cyclones (Ankur et al.

2020). For the north Indian Ocean, in a 2°C warming scenario, along with the intensity of cyclones, the rainfall rate associated with the cyclones is also projected to increase by ~18% (Knutson et al. 2020). A similar projected increase in the rainfall rates for the north Indian Ocean cyclones is reported by Walsh et al. (2016). In a global warming scenario, along with the north Indian Ocean, the rainfall rate associated with cyclones in different ocean basins is also projected to increase (Knutson et al. 2010; Lin et al. 2015).

## 8. Concluding remarks and suggested leads for future research

We have provided a review of our current understanding of the ocean-atmosphere interaction in response to cyclones in the north Indian Ocean. Figure 10 and 11 highlights some of the key ocean-cyclone interaction processes that we have understood so far. A large number of studies have investigated the cyclone-ocean interactions in the region but there are still several limitations to our understanding of these complex interactions.

### 8.1. Scope for improvement in our understanding of ocean-cyclone interactions

There is an in-depth understanding on the processes controlling the impact of ocean temperatures on the cyclones and the cyclone induced SST cooling in the two basins during the two cyclone seasons. However, a detailed quantification on the processes controlling the sub-surface temperatures and a statistical assessment of the recovery time of the SSTs and the ocean heat content need to be carried out. The pre-monsoon cyclone season is immediately followed by the monsoon season. Hence it is important to understand the mechanisms regulating the recovery time of ocean temperatures, as studies for other basins show that the cold wake induced by the cyclone can significantly modulate the ocean-atmosphere interaction and can reduce the cloud cover and rain for weeks after the passage of the cyclone (Ma et al. 2020; Pasquero et al. 2021). Chan and Chan (2012) show that the larger size of West Pacific cyclones is possibly associated with the higher SSTs, and the size of the cyclone influence the spatial extent of the cyclone induced cooling (Pun et al. 2018). We require a similar understanding for the north Indian Ocean that can provide insights into the role of pre-monsoon cyclone-ocean interactions on the subsequent monsoon season.

Cyclones can induce ocean heat transport out of the basin and can significantly alter the net heat content, as observed in the Bay of Bengal. However, a detailed analysis of the poleward ocean-heat transport in response to the cyclones is pending for the Arabian Sea. The ocean response to cyclone is not symmetrical along the track of cyclones, as discussed in section 4 (Figure 10). Figure 10 summarizes the asymmetrical ocean response to the cyclone due to the asymmetric wind forcing and ocean currents. There is scope for a comprehensive understanding that includes research on the evolution of the ocean surface and sub-surface currents in response to this asymmetric ocean-cyclone interaction. Cyclones in the Bay of Bengal are observed to intensify as they move across warm core eddies and weakens as they move across cold core eddies. It is important that we have a similar understanding of the role of ocean eddies in governing the intensity of cyclones in the Arabian Sea.

As mentioned in section 5, a lot of research have explored the biological response of the Arabian Sea and the Bay of Bengal in the two seasons. The spatial extent and magnitude of the chlorophyll blooms depend on the cyclone translation speed and intensity, and ocean subsurface conditions such as barrier layer, mixed layer and ocean stratification. This research is important since an increase or decrease in the number of cyclones can significantly affect the biological conditions of the north Indian Ocean, and that may have a significant influence on the marine food chain. More studies are required to comprehensively evaluate the contribution of cyclones in increasing the seasonal net-productivity in the two basins and the impact of size of the cyclone on these chlorophyll blooms.

8.2. Improving cyclone forecasts with better observations

The forecast of the cyclone tracks and landfall positions of the north Indian Ocean cyclones have significantly improved with time and human fatalities have reduced in the recent period (Mohapatra et al. 2013b, 2015; Ray et al. 2021). However, the forecast of the intensity of the cyclones with high lead time remains a challenge, particularly when the intensification is rapid (Mohapatra et al. 2013a). This is due to the gaps in in-situ ocean observations which lead to poor initial condition in the numerical weather prediction models resulting in an increase in forecast error of these models (Mohapatra et al. 2013a; Bell et al. 2020). Webster (2013) points out that the cost of extended 2-week forecasts for South Asia for hydrometeorological hazards including tropical cyclones is relatively small, amounting to a maximum of $3 million per year. Considering that the economic losses due to these disasters exceed $3 billion per year (International Disaster Data Base) and continues to rise with global warming, there is an urgent need to improve the extended predictability of these cyclones. We have seen that the ocean plays a very important role in modulating the cyclone genesis and its intensity. Hence implementing high-resolution surface and subsurface ocean observations instruments for closely monitoring the changes in the characteristics of the ocean is a priority (Beal et al. 2020). An increase in the number of moorings (e.g., RAMA, and Ocean Moored buoy Network for the Northern Indian Ocean (OMNI)) and buoys (ARGO) for the profiling of the oceanic variables in real-time can give us more accurate estimates of the ocean. Improved ocean observations play a crucial role in validating and initializing ocean-atmosphere coupled models and are assimilated into regional and global models which can play a crucial role in cyclone forecasting. The number of tide gauges installed at the coastal boundaries of the Indian Ocean rim countries needs to be increased, as it will give a better idea about the storm surges caused by the cyclones along the coastal belt. While forecasting the intensity of cyclones, all these oceanic conditions and associated interactions need to be tightly integrated into the model.

The heat flux exchange between the ocean and the atmosphere is one of the major factors that control the intensity of the cyclone. The estimation of heat fluxes at the ocean surface remains a challenge (Yu 2019). The flux measurements from the RAMA moored buoy array are essential in improving the flux estimates during the genesis of cyclones which can significantly improve the cyclone forecast (Beal et al. 2020). As shown earlier, the salinity plays a crucial role in governing the ocean characteristics and the ocean-cyclone interaction. Therefore, the in-situ observations of surface salinity during the cyclone also need to be improved as satellites fail to sense the salinity variations of the upper ocean due to deep convective clouds during the time of cyclone. Currently, there are only two satellite scatterometers that provide approximately 60% coverage of the ocean at 6-hourly interval, the de-correlation time scale of the diurnal cycle (Beal et al. 2019). The spatial and temporal coverage of satellite-derived winds and wind stress estimates over the open ocean needs to be increased especially at the time of a cyclone as it will give a better idea about the ocean-atmosphere coupling and diurnal variations during the cyclone.

8.3. Understanding the changes in ocean-cyclone interactions under rapid ocean warming

The north Indian Ocean is warming at a rapid pace. As a result of ocean warming, the TCHP is increasing at a rate of 0.53 kJ cm$^{-2}$ in the Arabian Sea (Rajeevan et al. 2013). This highlights that the ocean is becoming more conducive to fuel intense cyclones. A large number (62%) of intense cyclones (wind speed >100 knots) in the north Indian Ocean attain their maximum intensity within 200 km away from the coast (Hoarau et al. 2012). Hence, further studies should not only focus on the changes in the open ocean characteristics but also on the changes near the coast and its role in the increasing intensity of cyclones, especially in the Arabian Sea. The cyclone size and cyclone induced rain rate are projected to increase with the increase in the SSTs in a global warming scenario (Lin et al. 2015; Chavas et al. 2016; Sun et al. 2017). Hence, it is important to understand the role of SSTs on the cyclone size, the two-way feedback (that is the impact of size of cyclone on the SSTs) and the associated rain in the north Indian Ocean, as an increase in cyclone size and the rain rate may broaden the cyclone impacted areas and will enhance the risk of flash floods in the coastal areas. As climate models continue to project continued rapid warming of the north Indian Ocean (Ogata et al. 2014), more studies should focus on how the ocean-atmospheric processes during cyclones will change in the future.

**Acknowledgements**

We thank Juby Aleyas Koll for preparing the schematic illustration for Figure 10. We thank the anonymous reviewers for the useful suggestions.

# References

Akhil VP, Durand F, Lengaigne M, et al (2014) A modeling study of the processes of surface salinity seasonal cycle in the Bay of Bengal. J Geophys Res Ocean 119:3926–3947. doi: https://doi.org/10.1002/2013JC009632

Alam MM, Hossain MA, Shafee S (2003) Frequency of Bay of Bengal cyclonic storms and depressions crossing different coastal zones. Int J Climatol 23:1119–1125. doi: 10.1002/joc.927

Albert J, Bhaskaran PK (2020) Ocean heat content and its role in tropical cyclogenesis for the Bay of Bengal basin. Clim Dyn 55:3343–3362. doi: 10.1007/s00382-020-05450-9

Ali A (1999) Climate change impacts and adaptation assessment in Bangladesh. Clim Res 12:

Ali MM, Swain D, Kashyap T, et al (2013) Relationship between cyclone intensities and sea surface temperature in the tropical Indian Ocean. IEEE Geosci Remote Sens Lett 10:841–844. doi: 10.1109/LGRS.2012.2226138

Ali SA, Mao Z, Wu J, et al (2020) Satellite Evidence of Upper Ocean Responses to Cyclone Nilofar. Atmosphere-Ocean 58:13–24. doi: 10.1080/07055900.2019.1700097

Anandh TS, Das BK, Kumar B, et al (2018) Analyses of the oceanic heat content during 1980–2014 and satellite-era cyclones over Bay of Bengal. Int J Climatol 38:5619–5632. doi: 10.1002/joc.5767

Anandh TS, Das BK, Kuttippurath J, Chakraborty A (2020) A coupled model analyses on the interaction between oceanic eddies and tropical cyclones over the Bay of Bengal. Ocean Dyn 70:327–337. doi: 10.1007/s10236-019-01330-x

Ankur K, Busireddy NKR, Osuri KK, Niyogi D (2020) On the relationship between intensity changes and rainfall distribution in tropical cyclones over the North Indian Ocean. Int J Climatol 40:2015–2025. doi: 10.1002/joc.6315

Babin SM, Carton JA, Dickey TD, Wiggert JD (2004) Satellite evidence of hurricane-induced phytoplankton blooms in an oceanic desert. J. Geophys. Res. Ocean. 109

Babu KN, Sharma R, Agarwal N, et al (2004) Study of the mixed layer depth variations within the north Indian Ocean using a 1-D model. J Geophys Res Ocean 109:. doi: 10.1029/2003JC002024

Badarinath KVS, Kharol SK, Dileep PK, Prasad VK (2009) Satellite observations on cyclone-induced upper ocean cooling and modulation of surface winds-A study on tropical ocean region. IEEE Geosci Remote Sens Lett 6:481–485. doi: 10.1109/LGRS.2009.2018487

Balaguru K, Chang P, Saravanan R, et al (2012) Ocean barrier layers' effect on tropical cyclone intensification. Proc Natl Acad Sci 109:14343–14347. doi: 10.1073/pnas.1201364109

Balaguru K, Ruby Leung L, Yoon JH (2013) Oceanic control of Northeast Pacific hurricane activity at interannual timescales. Environ Res Lett 8:44009. doi: 10.1088/1748-9326/8/4/044009

Balaguru K, Taraphdar S, Leung LR, Foltz GR (2014) Increase in the intensity of postmonsoon Bay of Bengal tropical cyclones. Geophys Res Lett 41:3594–3601. doi: 10.1002/2014GL060197

Balaji M, Chakraborty A, Mandal M (2018) Changes in tropical cyclone activity in north Indian Ocean during satellite era (1981–2014). Int J Climatol 38:2819–2837. doi: 10.1002/joc.5463

Bao JW, Wilczak JM, Choi JK, Kantha LH (2000) Numerical simulations of air-sea interaction under high wind conditions using a coupled model: A study of Hurricane development. Mon Weather Rev 128:2190–2210. doi: 10.1175/1520-0493(2000)128<2190:NSOASI>2.0.CO;2

Beal LM, Vialard J, Roxy MK, et al (2020) A roadmap to IndOOS-2: Better observations of the rapidly-warming Indian Ocean. Bull Am Meteorol Soc 1–50. doi: 10.1175/bams-d-19-0209.1

Beal L. M., J. Vialard, M. K. Roxy and lead authors (2019) IndOOS-2: A roadmap to sustained observations of the Indian Ocean for 2020-2030, CLIVAR-4/2019, GOOS-237, 206 pp. doi: 10.36071/clivar.rp.4.2019

Behera SK, Deo AA, Salvekar PS (1998) Investigation of mixed layer response to Bay of Bengal cyclone using a simple ocean model. Meteorol Atmos Phys 65:77–91. doi: 10.1007/BF01030270

Behera SK, Salvekar PS, Yamagata T (2000) Simulation of interannual SST variability in the tropical Indian Ocean. J Clim 13:3487–3499. doi: 10.1175/1520-0442(2000)013<3487:SOISVI>2.0.CO;2

Bell SS, Chand SS, Tory KJ, et al (2020) North Indian Ocean tropical cyclone activity in CMIP5 experiments: Future projections using a model-independent detection and tracking scheme. Int J Climatol 1–14. doi: 10.1002/joc.6594

Bender MA, Ginis I (2000) Real-case simulations of hurricane-ocean interaction using a high-resolution coupled model: Effects on hurricane intensity. Mon Weather Rev 128:917–946. doi: 10.1175/1520-0493(2000)128<0917:RCSOHO>2.0.CO;2

Bender MA, Ginis I, Kurihara Y (1993) Numerical simulations of tropical cyclone-ocean interaction with a high-resolution coupled model. J Geophys Res 98:245–268. doi: 10.1029/93jd02370

Bhardwaj P, Pattanaik DR, Singh O (2019) Tropical cyclone activity over Bay of Bengal in relation to El Niño-Southern Oscillation. Int J Climatol 39:5452–5469. doi: 10.1002/joc.6165

Bhatia K, Vecchi G, Murakami H, et al (2018) Projected Response of Tropical Cyclone Intensity and Intensification in a Global Climate Model. J Clim 31:8281–8303. doi: 10.1175/JCLI-D-17-0898.1

Bhattathiri PMA, Pant A, Sawant S, et al (1996) Phytoplankton production and chlorophyll distribution in the eastern and central Arabian Sea in 1994-1995. Curr Sci 71:857–862

Bond NA, Cronin MF, Sabine C, et al (2011) Upper ocean response to Typhoon Choi-Wan as measured by the Kuroshio Extension Observatory mooring. J Geophys Res Ocean 116:

Bongirwar V, Rakesh V, Kishtawal CM, Joshi PC (2011) Impact of satellite observed microwave

SST on the simulation of tropical cyclones. Nat Hazards 58:929–944. doi: 10.1007/s11069-010-9699-y

Brooks DA (1983) The wake of Hurricane Allen in the western Gulf of Mexico. J. Phys. Ocean. 13:117–129

Busireddy NKR, Ankur K, Osuri KK, et al (2019) The response of ocean parameters to tropical cyclones in the Bay of Bengal. Q J R Meteorol Soc 145:3320–3332. doi: 10.1002/qj.3622

Byju P, Kumar SP (2011) Physical and biological response of the Arabian Sea to tropical cyclone Phyan and its implications. Mar Environ Res 71:325–330. doi: 10.1016/j.marenvres.2011.02.008

Byju P, Prasanna Kumar S (2011) Physical and biological response of the Arabian Sea to tropical cyclone Phyan and its implications. Mar Environ Res 71:325–330. doi: 10.1016/j.marenvres.2011.02.008

Camargo SJ, Sobel AH (2005) Western North Pacific tropical cyclone intensity and ENSO. J Clim 18:2996–3006. doi: 10.1175/JCLI3457.1

Camp J, Roberts M, Maclachlan C, et al (2015) Seasonal forecasting of tropical storms using the Met Office GloSea5 seasonal forecast system. Q J R Meteorol Soc 141:2206–2219. doi: 10.1002/qj.2516

Cecil DJ (2007) Satellite-derived rain rates in vertically sheared tropical cyclones. Geophys Res Lett 34:1–4. doi: 10.1029/2006GL027942

Chacko N (2018a) Effect of Cyclone Thane in the Bay of Bengal Explored Using Moored Buoy Observations and Multi-platform Satellite Data. J Indian Soc Remote Sens 46:821–828. doi: 10.1007/s12524-017-0748-9

Chacko N (2018b) Insights into the haline variability induced by cyclone Vardah in the Bay of Bengal using SMAP salinity observations. Remote Sens Lett 9:1205–1213. doi: 10.1080/2150704X.2018.1519271

Chacko N (2017) Chlorophyll bloom in response to tropical cyclone Hudhud in the Bay of Bengal: Bio-Argo subsurface observations. Deep Res Part I Oceanogr Res Pap 124:66–72. doi: 10.1016/j.dsr.2017.04.010

Chacko N (2019) Differential chlorophyll blooms induced by tropical cyclones and their relation to cyclone characteristics and ocean pre-conditions in the Indian Ocean. J Earth Syst Sci 128:. doi: 10.1007/s12040-019-1207-5

Chaitanya AVS, Lengaigne M, Vialard J, et al (2014) Salinity measurements collected by fishermen reveal a "river in the sea" flowing along the eastern coast of India. Bull Am Meteorol Soc 95:1897–1908. doi: 10.1175/BAMS-D-12-00243.1

Chan KTF, Chan JCL (2012) Size and strength of tropical cyclones as inferred from QuikSCAT data. Mon Weather Rev 140:811–824. doi: 10.1175/MWR-D-10-05062.1

Chaudhuri D, Sengupta D, D'Asaro E, et al (2019) Response of the salinity-stratified Bay of Bengal to cyclone Phailin. J Phys Oceanogr 49:1121–1140. doi: 10.1175/JPO-D-18-0051.1

Chauhan A, Singh R, Kumar R (2021) Dynamics of Amphan Cyclone and associated changes in ocean , land meteorological and atmospheric parameters. Mausam 72:215–228

Chavas DR, Lin N, Dong W, Lin Y (2016) Observed tropical cyclone size revisited. J Clim 29:2923–2939. doi: 10.1175/JCLI-D-15-0731.1

Chen S, Campbell TJ, Jin H, et al (2010) Effect of two-way air-sea coupling in high and low wind speed regimes. Mon Weather Rev 138:3579–3602. doi: 10.1175/2009MWR3119.1

Chen Y, Zhang F, Green BW, Yu X (2018) Impacts of ocean cooling and reduced wind drag on Hurricane Katrina (2005) based on numerical simulations. Mon Weather Rev 146:287–306. doi: 10.1175/MWR-D-17-0170.1

Cheng L, Trenberth KE, Fasullo J, et al (2017) Improved estimates of ocean heat content from 1960 to 2015. Sci Adv 3:1–11. doi: 10.1126/sciadv.1601545

Chowdhury RR, Prasanna Kumar S, Chakraborty A (2020a) A study on the physical and biogeochemical responses of the Bay of Bengal due to cyclone Madi. J Oper Oceanogr 1–22

Chowdhury RR, Prasanna Kumar S, Narvekar J, Chakraborty A (2020b) Back-to-Back Occurrence of Tropical Cyclones in the Arabian Sea During October–November 2015: Causes and Responses. J Geophys Res Ocean 125:1–23. doi: 10.1029/2019JC015836

Cione JJ, Uhlhorn EW (2003) Sea surface temperature variability in hurricanes: Implications with respect to intensity change. Mon Weather Rev 131:1783–1796. doi: 10.1175//2562.1

D'Asaro EA, Sanford TB, Niiler PP, Terrill EJ (2007) Cold wake of Hurricane Frances. Geophys Res Lett 34:2–7. doi: 10.1029/2007GL030160

Dare RA, Mcbride JL (2011) Sea surface temperature response to tropical cyclones. Mon Weather Rev 139:3798–3808. doi: 10.1175/MWR-D-10-05019.1

de Boyer Montégut C, Mignot J, Lazar A, Cravatte S (2007) Control of salinity on the mixed layer depth in the world ocean: 1. General description. J Geophys Res Ocean 112:. doi: 10.1029/2006JC003953

DeMaria M, Kaplan J (1999) An updated Statistical Hurricane Intensity Prediction Scheme (SHIPS) for the Atlantic and eastern North Pacific basins. Weather Forecast 14:326–337. doi: 10.1175/1520-0434(1999)014<0326:AUSHIP>2.0.CO;2

Deshpande M, Singh VK, Kranthi GM, et al (2021) Changing Status of Tropical Cyclones over the North Indian Ocean. Clim Dyn 1–23. doi: https://doi.org/10.1007/s00382-021-05880-z

Emanuel K (2005) Increasing destructiveness of tropical cyclones over the past 30 years. Nature 436:686–688. doi: 10.1038/nature03906

Emanuel K (2001) Contribution of tropical cyclones to meridional heat transport by the oceans. J Geophys Res Atmos 106:14771–14781. doi: 10.1029/2000JD900641

Emanuel KA (1986) An air-sea interaction theory for tropical cyclones. Part I: steady-state maintenance. J. Atmos. Sci. 43:585–604

Evan AT, Kossin JP, Chung CE, Ramanathan V (2011) Arabian Sea tropical cyclones intensified

by emissions of black carbon and other aerosols. Nature 479:94–97. doi: 10.1038/nature10552

Fan K, Wang X, He Z (2020) Control of salinity stratification on recent increase in tropical cyclone intensification rates over the postmonsoon Bay of Bengal. Environ Res Lett 15:. doi: 10.1088/1748-9326/ab9690

Felton CS, Subrahmanyam B, Murty VSN (2013) ENSO-modulated cyclogenesis over the Bay of Bengal. J Clim 26:9806–9818. doi: 10.1175/JCLI-D-13-00134.1

Francis PA, Gadgil S, Vinayachandran PN (2007) Triggering of the positive Indian Ocean dipole events by severe cyclones over the Bay of Bengal. Tellus, Ser A Dyn Meteorol Oceanogr 59 A:461–475. doi: 10.1111/j.1600-0870.2007.00254.x

Fudeyasu H, Ito K, Miyamoto Y (2018) Characteristics of tropical cyclone rapid intensification over the Western North Pacific. J Clim 31:8917–8930. doi: 10.1175/JCLI-D-17-0653.1

Gao S, Chiu LS (2010) Surface latent heat flux and rainfall associated with rapidly intensifying tropical cyclones over the western North Pacific. Int J Remote Sens 31:4699–4710. doi: 10.1080/01431161.2010.485149

Gao S, Jia S, Wan Y, et al (2019) The role of latent heat flux in tropical cyclogenesis over the Western North Pacific: Comparison of developing versus non-developing disturbances. J Mar Sci Eng 7:. doi: 10.3390/jmse7020028

Gao S, Zhai S, Chiu LS, Xia D (2016) Satellite air-sea enthalpy flux and intensity change of tropical cyclones over the western North Pacific. J Appl Meteorol Climatol 55:425–444. doi: 10.1175/JAMC-D-15-0171.1

Ghetiya S, Nayak RK (2020) Genesis potential parameter using satellite derived daily tropical cyclone heat potential for North Indian ocean. Int J Remote Sens 41:8932–8945. doi: 10.1080/01431161.2020.1795299

Girishkumar MS, Ravichandran M (2012) The influences of ENSO on tropical cyclone activity in the Bay of Bengal during October-December. J Geophys Res Ocean 117:1–13. doi: 10.1029/2011JC007417

Girishkumar MS, Suprit K, Chiranjivi J, et al (2014) Observed oceanic response to tropical cyclone Jal from a moored buoy in the south-western Bay of Bengal. Ocean Dyn 64:325–335. doi: 10.1007/s10236-014-0689-6

Girishkumar MS, Suprit K, Vishnu S, et al (2015) The role of ENSO and MJO on rapid intensification of tropical cyclones in the Bay of Bengal during October–December. Theor Appl Climatol 120:797–810. doi: 10.1007/s00704-014-1214-z

Gopalan AKS, Gopala Krishna V V., Ali MM, Sharma R (2000) Detection of Bay of Bengal eddies from TOPEX and in situ observations. J Mar Res 58:721–734. doi: 10.1357/002224000321358873

Gopalkrishna VV, Murty VSN, Sarma MSS, Sastry JS (1993) Thermal response of upper layeys of Bay of Bengal to forcing of a severe cyclonic storm: A case study. Indian J Geo-Marine Sci 22:8–11

Hallam S, Guishard M, Josey SA, et al (2021) Increasing tropical cyclone intensity and potential intensity in the subtropical Atlantic around Bermuda from an ocean heat content perspective 1955-2019. Environ. Res. Lett. 16:34052

Hernandez O, Jouanno J, Durand F (2016) Do the Amazon and Orinoco freshwater plumes really matter for hurricane-induced ocean surface cooling? J Geophys Res Ocean 121:2119–2141. doi: 10.1002/2015JC011021

Hoarau K, Bernard J, Chalonge L (2012) Intense tropical cyclone activities in the northern Indian Ocean. Int J Climatol 32:1935–1945. doi: 10.1002/joc.2406

Hsu PC, Ho CR (2019) Typhoon-induced ocean subsurface variations from glider data in the Kuroshio region adjacent to Taiwan. J Oceanogr 75:1–21. doi: 10.1007/s10872-018-0480-2

Huang P, Sanford TB, Imberger J (2009) Heat and turbulent kinetic energy budgets for surface layer cooling induced by the passage of Hurricane Frances (2004). J Geophys Res Ocean 114:1–14. doi: 10.1029/2009JC005603

Jacob SD, Koblinsky CJ (2007) Effects of precipitation on the upper-ocean response to a hurricane. Mon Weather Rev 135:2207–2225. doi: 10.1175/MWR3366.1

Jacob SD, Shay LK, Mariano AJ, Black PG (2000) The 3D oceanic mixed layer response to Hurricane Gilbert. J Phys Oceanogr 30:1407–1429. doi: 10.1175/1520-0485(2000)030<1407:TOMLRT>2.0.CO;2

Jaimes B, Shay LK, Uhlhorn EW (2015) Enthalpy and momentum fluxes during hurricane earl relative to underlying ocean features. Mon Weather Rev 143:111–131. doi: 10.1175/MWR-D-13-00277.1

Jangir B, Swain D, Ghose SK (2020) Influence of eddies and tropical cyclone heat potential on intensity changes of tropical cyclones in the North Indian Ocean. Adv Sp Res 1–14. doi: 10.1016/j.asr.2020.01.011

Jansen MF, Ferrari R, Mooring TA (2010) Seasonal versus permanent thermocline warming by tropical cyclones. Geophys Res Lett 37:n/a-n/a. doi: 10.1029/2009GL041808

Jarosz E, Hallock ZR, Teague WJ (2007) Near-inertial currents in the DeSoto Canyon region. Cont Shelf Res 27:2407–2426. doi: 10.1016/j.csr.2007.06.014

Joseph JK, Balchand AN, Hareeshkumar P V., Rajesh G (2007) Inertial oscillation forced by the September 1997 cyclone in the Bay of Bengal. Curr Sci 92:790–794

Jourdain NC, Lengaigne M, Vialard JR, et al (2013) Observation-based estimates of surface cooling inhibition by heavy rainfall under tropical cyclones. J Phys Oceanogr 43:205–221. doi: 10.1175/JPO-D-12-085.1

Jullien S, Marchesiello P, Menkes CE, et al (2014) Ocean feedback to tropical cyclones: climatology and processes. Clim Dyn 43:2831–2854. doi: 10.1007/s00382-014-2096-6

Jullien S, Menkes CE, Marchesiello P, et al (2012) Impact of tropical cyclones on the heat budget of the South Pacific Ocean. J Phys Oceanogr 42:1882–1906. doi: 10.1175/JPO-D-11-0133.1

Kaplan J, DeMaria M (2003) Large-scale characteristics of rapidly intensifying tropical cyclones

in the North Atlantic basin. Weather Forecast 18:1093–1108. doi: 10.1175/1520-0434(2003)018<1093:LCORIT>2.0.CO;2

Kikuchi K, Wang B, Fudeyasu H (2009) Genesis of tropical cyclone Nargis revealed by multiple satellite observations. Geophys Res Lett 36:1–5. doi: 10.1029/2009GL037296

Kim HS, Lozano C, Tallapragada V, et al (2014) Performance of ocean simulations in the coupled HWRF-HYCOM model. J Atmos Ocean Technol 31:545–559. doi: 10.1175/JTECH-D-13-00013.1

Knapp KR, Kruk MC, Levinson DH, et al (2010) The International Best Track Archive for Climate Stewardship (IBTrACS). Bull Am Meteorol Soc 91:363–376. doi: 10.1175/2009BAMS2755.1

Knutson T, Camargo SJ, Chan JCL, et al (2020) Tropical cyclones and climate change assessment: Part2: projected response to anthropogenic warming. Bull Am Meteorol Soc 101:E303–E322. doi: 10.1175/BAMS-D-18-0189.1

Knutson TR, McBride JL, Chan J, et al (2010) Tropical cyclones and climate change. Nat Geosci 3:157–163. doi: 10.1038/ngeo779

Kotal SD, Kundu PD, Bhowmik SKR (2009) Analysis of cyclogeneis parameter for developing and nondeveloping low-pressure systems over the Indian Sea. Nat Hazards 50:389–402. doi: 10.1007/s11069-009-9348-5

Krishna KM (2009) Intensifying tropical cyclones over the North Indian Ocean during summer monsoon-Global warming. Glob Planet Change 65:12–16. doi: 10.1016/j.gloplacha.2008.10.007

Krishna KM, Rao SR (2009) Study of the intensity of super cyclonic storm GONU using satellite observations. Int J Appl Earth Obs Geoinf 11:108–113. doi: 10.1016/j.jag.2008.11.001

Krishnamohan KS, Mohanakumar K, Joseph P V. (2012) The influence of Madden-Julian Oscillation in the genesis of North Indian Ocean tropical cyclones. Theor Appl Climatol 109:271–282. doi: 10.1007/s00704-011-0582-x

Kumar BP, D'Asaro E, Suresh kumar N, Ravichandran M (2019) Widespread cooling of the Bay of Bengal by tropical storm Roanu. Deep Res Part II Top Stud Oceanogr 168:104652. doi: 10.1016/j.dsr2.2019.104652

Kundu PK (1976) An analysis of inertial oscillations observed near Oregon coast. J Phys Oceanogr 6:879–893

Kurien P, Ikeda M, Valsala VK (2010) Mesoscale variability along the east coast of India in spring as revealed from satellite data and OGCM simulations. J Oceanogr 66:273–289. doi: 10.1007/s10872-010-0024-x

Kuttippurath J, Sunanda N, Martin M V., Chakraborty K (2021) Tropical storms trigger phytoplankton blooms in the deserts of north Indian Ocean. npj Clim Atmos Sci 4:1–12. doi: 10.1038/s41612-021-00166-x

Latha TP, Rao KH, Nagamani P V, et al (2015) Impact of Cyclone PHAILIN on Chlorophyll- a

Concentration and Productivity in the Bay of Bengal. Int J Geosci 6:473–480

Leaman KD, Sanford TB (1975) Energy propagation of inertial waves. J Geophys Res 80:1975–1978

Lee SK, Park W, Baringer MO, et al (2015) Pacific origin of the abrupt increase in Indian Ocean heat content during the warming hiatus. Nat Geosci 8:445–449. doi: 10.1038/NGEO2438

Leipper DF, Volgenau LD (1972) Hurricane heat potential of the Gulf of Mexico. J Phys Oceanogr 2(3):218–224

Lengaigne M, Guillaume SN, Jérôme S, et al (2018) Influence of air – sea coupling on Indian Ocean tropical cyclones. Clim Dyn 0:0. doi: 10.1007/s00382-018-4152-0

Li Y (2017) Bay of Bengal salinity stratification and Indian summermonsoon intraseasonal oscillation: 1. Intraseasonal variabilityand causes. J Geophys Res Ocean 122:2647–2651. doi: 10.1002/2017JC012961.Received

Li Z, Li T, Yu W, et al (2016) What controls the interannual variation of tropical cyclone genesis frequency over Bay of Bengal in the post-monsoon peak season? Atmos Sci Lett 17:148–154. doi: 10.1002/asl.636

Li Z, Yu W, Li T, et al (2013) Bimodal character of cyclone climatology in the Bay of Bengal modulated by monsoon seasonal cycle. J Clim 26:1033–1046. doi: 10.1175/JCLI-D-11-00627.1

Liebmann B, Hendon H, Glick D (1994) The Relationship between Tropical Cyclones of the Western Pacific and Indian Oceans and the Madden-Julian Oscillation. J Meteorol Soc Japan Ser // 72:401–411

Lin II, Pun IF, Lien CC (2014) "category-6" supertyphoon Haiyan in global warming hiatus: Contribution from subsurface ocean warming. Geophys Res Lett 41:8547–8553. doi: 10.1002/2014GL061281

Lin II, Pun IF, Wu CC (2009) Upper-ocean thermal structure and the western north pacific category 5 typhoons. Part II: Dependence on translation speed. Mon Weather Rev 137:3744–3757. doi: 10.1175/2009MWR2713.1

Lin Y, Zhao M, Zhang M (2015) Tropical cyclone rainfall area controlled by relative sea surface temperature. Nat Commun 6:1–7. doi: 10.1038/ncomms7591

Lin YC, Oey LY (2016) Rainfall-enhanced blooming in typhoon wakes. Sci Rep 6:1–9. doi: 10.1038/srep31310

Liu B, Liu H, Xie L, et al (2011) A Coupled atmosphere-wave-ocean modeling system: simulation of the intensity of an idealized tropical cyclone. Mon Weather Rev 139:132–152. doi: 10.1175/2010MWR3396.1

Liu B, Zhu C (2020) Boosting Effect of Tropical Cyclone "Fani" on the Onset of the South China Sea Summer Monsoon in 2019. J Geophys Res Atmos 125:1–16. doi: 10.1029/2019JD031891

Liu F, Zhang H, Ming J, et al (2020) Importance of precipitation on the upper ocean salinity

response to typhoon kalmaegi (2014). Water 12:614. doi: 10.3390/w12020614

Liu Z, Xu J, Zhu B, et al (2007) The upper ocean response to tropical cyclones in the northwestern Pacific analyzed with Argo data. Chinese J Oceanol Limnol 25:123–131. doi: 10.1007/s00343-007-0123-8

Lloyd ID, Vecchi GA (2011) Observational evidence for oceanic controls on hurricane intensity. J Clim 24:1138–1153. doi: 10.1175/2010JCLI3763.1

Lü H, Zhao X, Sun J, et al (2020) A case study of a phytoplankton bloom triggered by a tropical cyclone and cyclonic eddies. PLoS One 15:1–18. doi: 10.1371/journal.pone.0230394

Ma Z, Fei J, Lin Y, Huang X (2020) Modulation of Clouds and Rainfall by Tropical Cyclone's Cold Wakes. Geophys Res Lett 47:1–8. doi: 10.1029/2020GL088873

Ma Z, Fei J, Liu L, et al (2013) Effects of the cold core eddy on tropical cyclone intensity and structure under idealized air-sea interaction conditions. Mon Weather Rev 141:1285–1303. doi: 10.1175/MWR-D-12-00123.1

Madden RA, Julian PR (1971) Detection of a 40–50 day oscillation in the zonal wind in the tropical Pacific. J Atmos Sci 28:702–708. doi: 10.1175/1520-0469(1971)028<0702:DOADOI>2.0.CO;2

Madhupratap M, Prasanna Kumar S, Bhattathiri PMA, et al (1996) Mechanism of the biological response to winter cooling in the northeastern Arabian Sea. Nature 384:549–552. doi: 10.1038/384549a0

Madsen H, Jakobsen F (2004) Cyclone induced storm surge and flood forecasting in the northern Bay of Bengal. Coast Eng 51:277–296. doi: 10.1016/j.coastaleng.2004.03.001

Mahala BK, Nayak BK, Mohanty PK (2015) Impacts of ENSO and IOD on tropical cyclone activity in the Bay of Bengal. Nat Hazards 75:1105–1125. doi: 10.1007/s11069-014-1360-8

Mahapatra DK, Rao AD, Babu S V., Srinivas C (2007) Influence of coast line on upper ocean's response to the tropical cyclone. Geophys Res Lett 34:9–11. doi: 10.1029/2007GL030410

Mandal M, Mohanty UC, Sinha P, Ali MM (2007) Impact of sea surface temperature in modulating movement and intensity of tropical cyclones. Nat Hazards 41:413–427. doi: 10.1007/s11069-006-9051-8

Mandal S, Sil S, Shee A, Venkatesan R (2018) Upper ocean and subsurface variability in the Bay of Bengal During Cyclone Roanu: a synergistic view using in situ and satellite observations. Pure Appl Geophys 175:4605–4624. doi: 10.1007/s00024-018-1932-8

Maneesha K, Murty VSN, Ravichandran M, et al (2012) Upper ocean variability in the Bay of Bengal during the tropical cyclones Nargis and Laila. Prog Oceanogr 106:49–61. doi: 10.1016/j.pocean.2012.06.006

Maneesha K, Prasad DH, Patnaik KVKRK (2019) Biophysical responses to tropical cyclone Hudhud over the Bay of Bengal. J Oper Oceanogr 0:1–11. doi: 10.1080/1755876X.2019.1684135

Maneesha K, Sadhuram Y, Prasad KVSR (2015) Role of upper ocean parameters in the genesis ,

intensification and tracks of cyclones over the Bay of Bengal. J Oper Oceanogr 8:133–146. doi: 10.1080/1755876X.2015.1087185

Mathew S, Natesan U, Latha G, et al (2018) Observed warming of sea surface temperature in response to tropical cyclone Thane in the Bay of Bengal. Curr Sci 114:1407–1413. doi: 10.18520/cs/v114/i07/1407-1413

McPhaden MJ, Foltz GR, Lee T, et al (2009) Ocean-Atmosphere Interactions During Cyclone Nargis. Eos, Trans Am Geophys Union 90:53–54

Mei W, Lien CC, Lin II, Xie SP (2015) Tropical cyclone-induced ocean response: A comparative study of the south China sea and tropical northwest Pacific. J Clim 28:5952–5968. doi: 10.1175/JCLI-D-14-00651.1

Mei W, Pasquero C (2013) Spatial and temporal characterization of sea surface temperature response to tropical cyclones. J Clim 26:3745–3765. doi: 10.1175/JCLI-D-12-00125.1

Mei W, Pasquero C, Primeau F (2012) The effect of translation speed upon the intensity of tropical cyclones over the tropical ocean. Geophys Res Lett 39:1–6. doi: 10.1029/2011GL050765

Mei W, Xie SP (2016) Intensification of landfalling typhoons over the northwest Pacific since the late 1970s. Nat Geosci 9:753–757. doi: 10.1038/ngeo2792

Menkes CE, Lengaigne M, Levy M, et al (2016) Global impact of tropical cyclones on primary production. Global Biogeochem Cycles 30:767–786. doi: 10.1002/2015GB005214.Received

Mohan GM, Srinivas C V., Naidu C V., et al (2015) Real-time numerical simulation of tropical cyclone Nilam with WRF: experiments with different initial conditions, 3D-Var and Ocean Mixed Layer Model. Nat Hazards 77:597–624. doi: 10.1007/s11069-015-1611-3

Mohanty S, Nadimpalli R, Osuri KK, et al (2019) Role of Sea Surface Temperature in Modulating Life Cycle of Tropical Cyclones over Bay of Bengal. Trop Cyclone Res Rev 8:68–83. doi: 10.1016/j.tcrr.2019.07.007

Mohanty UC, Osuri KK, Pattanayak S, Sinha P (2012) An observational perspective on tropical cyclone activity over Indian seas in a warming environment. Nat Hazards 63:1319–1335. doi: 10.1007/s11069-011-9810-z

Mohapatra M, Bandyopadhyay BK, Nayak DP (2013a) Evaluation of operational tropical cyclone intensity forecasts over north Indian Ocean issued by India Meteorological Department. Nat Hazards 68:433–451. doi: 10.1007/s11069-013-0624-z

Mohapatra M, Nayak DP, Sharma M, et al (2013b) Evaluation of official tropical cyclone track forecast issued by India Meteorological Department. J Earth Syst Sci 122:589–601. doi: 10.1007/s12040-015-0581-x

Mohapatra M, Nayak DP, Sharma M, et al (2015) Evaluation of official tropical cyclone landfall forecast issued by India Meteorological Department. J Earth Syst Sci 124:861–874. doi: 10.1007/s12040-015-0581-x

Mohapatra M, Sharma M (2019) Cyclone warning services in India during recent years: A review. Mausam 70:635–666

Monaldo FM, Sikora TD, Babin SM, Sterner RE (1997) Satellite imagery of sea surface temperature cooling in the wake of Hurricane Edouard (1996). Mon Weather Rev 125:2716–2721. doi: 10.1175/1520-0493(1997)125<2716:SIOSST>2.0.CO;2

Mukherjee A, Shankar D, G Aparna S, et al (2013) Near-inertial currents off the east coast of India. Cont Shelf Res 55:29–39. doi: 10.1016/j.csr.2013.01.007

Muni Krishna K (2016) Observational study of upper ocean cooling due to Phet super cyclone in the Arabian Sea. Adv Sp Res 57:2115–2120. doi: 10.1016/j.asr.2016.02.024

Murakami H, Mizuta R, Shindo E (2012) Future changes in tropical cyclone activity projected by multi-physics and multi-SST ensemble experiments using the 60-km-mesh MRI-AGCM. Clim Dyn 39:2569–2584. doi: 10.1007/s00382-011-1223-x

Murakami H, Sugi M, Kitoh A (2013) Future Changes in Tropical Cyclone Activity in the North Indian Ocean Projected by the New High-Resolution MRI-AGCM. Clim Dyn 40:1949–1968. doi: 10.1007/978-94-007-7720-0_6

Murakami H, Vecchi GA, Underwood S (2017) Increasing frequency of extremely severe cyclonic storms over the Arabian Sea. Nat Clim Chang. doi: 10.1038/s41558-017-0008-6

Murty VSN, Rao DP, Sastry JS (1983) The lowering of sea surface temperature in the east central Arabian Sea associated with a cyclone. Mahasagar- Bull Natl Inst Oceanogr 16:67–71

Naik H, Naqvi SWA, Suresh T, Narvekar P V. (2008) Impact of a tropical cyclone on biogeochemistry of the central Arabian Sea. Global Biogeochem Cycles 22:1–11. doi: 10.1029/2007GB003028

Narvekar J, Prasanna Kumar S (2006) Seasonal variability of the mixed layer in the central Bay of Bengal and associated changes in nutrients and chlorophyll. Deep Res Part I Oceanogr Res Pap 53:820–835. doi: 10.1016/j.dsr.2006.01.012

Navaneeth KN, Martin M V., Joseph KJ, Venkatesan R (2019) Contrasting the upper ocean response to two intense cyclones in the Bay of Bengal. Deep Res Part I Oceanogr Res Pap 147:65–78. doi: 10.1016/j.dsr.2019.03.010

Neetu S, Lengaigne M, Vialard J, et al (2019) Premonsoon/Postmonsoon Bay of Bengal Tropical Cyclones Intensity: Role of Air-Sea Coupling and Large-Scale Background State. Geophys Res Lett 46:2149–2157. doi: 10.1029/2018GL081132

Neetu S, Lengaigne M, Vincent EM, et al (2012) Influence of upper-ocean stratification on tropical cyclone-induced surface cooling in the Bay of Bengal. J Geophys Res Ocean 117:1–19. doi: 10.1029/2012JC008433

Neumann CJ (1993) Global overview. Chapter 1, Global guide to tropical cyclone forecasting. Geneva

Ng EKW, Chan JCL (2012) Interannual variations of tropical cyclone activity over the north Indian Ocean. Int J Climatol 32:819–830. doi: 10.1002/joc.2304

Nigam T, Prakash KR, Pant V (2020) An assessment of the impact of oceanic initial conditions on the interaction of upper ocean with the tropical cyclones in the Arabian Sea. J Oper Oceanogr

13:121–137. doi: 10.1080/1755876X.2019.1658567

Ogata T, Mizuta R, Adachi Y, et al (2015) Effect of air-sea coupling on the frequency distribution of intense tropical cyclones over the northwestern Pacific. Geophys Res Lett 42:10415–10421. doi: 10.1002/2015GL066774

Ogata T, Ueda H, Inoue T, et al (2014) Projected future changes in the Asian monsoon: A comparison of CMIP3 and CMIP5 model results. J Meteorol Soc Japan 92:207–225. doi: 10.2151/jmsj.2014-302

Ooyama KV (1969) Numerical simulation of the life cycle of tropical cyclones. J Atmos Sci 26:3–40

Pasquero C, Desbiolles F, Meroni AN (2021) Air-Sea Interactions in the Cold Wakes of Tropical Cyclones. Geophys Res Lett 48:1–6. doi: 10.1029/2020GL091185

Pasquero C, Emanuel K (2008) Tropical cyclones and transient upper-ocean warning. J Clim 21:149–162. doi: 10.1175/2007JCLI1550.1

Patnaik KVKRK, Maneesha K, Sadhuram Y, et al (2014) East India Coastal Current induced eddies and their interaction with tropical storms over Bay of Bengal. J Oper Oceanogr 7:58–68. doi: 10.1080/1755876X.2014.11020153

Patra PK, Kumar MD, Mahowald N, Sarma VVSS (2007) Atmospheric deposition and surface stratification as controls of contrasting chlorophyll abundance in the North Indian Ocean. J Geophys Res Ocean 112:1–14. doi: 10.1029/2006JC003885

Piontkovski SA, Al-tarshi MH, Al-ismaili SM (2019) Inter-annual variability of mesoscale eddy occurrence in the western Arabian Sea. Int J Ocean Oceanogr 13:1–23

Pothapakula PK, Osuri KK, Pattanayak S, et al (2017) Observational perspective of SST changes during life cycle of tropical cyclones over Bay of Bengal. Nat Hazards 88:1769–1787. doi: 10.1007/s11069-017-2945-9

Prakash KR, Pant V (2017) Upper oceanic response to tropical cyclone Phailin in the Bay of Bengal using a coupled atmosphere-ocean model. Ocean Dyn 67:51–64. doi: 10.1007/s10236-016-1020-5

Prasad TG (2004) A comparison of mixed-layer dynamics between the Arabian Sea and Bay of Bengal: One-dimensional model results. J Geophys Res Ocean 109:. doi: 10.1029/2003jc002000

Prasad TG, Hogan PJ (2007) Upper-ocean response to Hurricane Ivan in a 1/25° nested Gulf of Mexico HYCOM. J Geophys Res Ocean 112:1–18. doi: 10.1029/2006JC003695

Prasanna Kumar S, Madhupratap M, Dileep Kumar M, et al (2000) Physical control of primary productivity on a seasonal scale in central and eastern Arabian Sea. Proc Indian Acad Sci Earth Planet Sci 109:433–441. doi: 10.1007/bf02708331

Premkumar K, Ravichandran M, Kalsi S.R., et al (2000) First result from a new observational system over the Indian Seas. Curr Sci 78:323–330

Price JF (1981) Upper Ocean Response to a Hurricane. J Phys Oceanogr 11:153–175. doi:

https://doi.org/10.1175/1520-0485(1981)011<0153:UORTAH>2.0.CO;2

Price JF, Sanford TB, Forristall GZ (1994) Forced stage response to a moving hurricane. J Phys Oceanogr 24:233–260

Pun IF, Lin II, Lien CC, Wu CC (2018) Influence of the size of Supertyphoon Megi (2010) on SST cooling. Mon Weather Rev 146:661–677. doi: 10.1175/MWR-D-17-0044.1

Qiu Y, Han W, Lin X, et al (2019) Upper-ocean response to the super tropical cyclone Phailin (2013) over the freshwater region of the Bay of Bengal. J Phys Oceanogr 49:1201–1228. doi: 10.1175/JPO-D-18-0228.1

Raghavan S, Varadarajan VM (1981) Radar estimate of rainfall and latent heat release in tropical cyclones of the Bay of Bengal. Mausam 32:247–252

Rajasree VPM, Kesarkar AP, Bhate JN, et al (2016) Appraisal of recent theories to understand cyclogenesis pathways of tropical cyclone Madi (2013). J Geophys Res Atmos 121:8949–8982. doi: 10.1038/175238c0

Rajeevan M, Srinivasan J, Niranjan Kumar K, et al (2013) On the epochal variation of intensity of tropical cyclones in the Arabian Sea. Atmos Sci Lett 14:249–255. doi: 10.1002/asl2.447

Rao AD (2007) Numerical Modeling of Cyclone's Impact on the Ocean—A Case Study of the Orissa Super Cyclone. J Coast Res 23:1245–1250

Rao AD, Joshi M, Jain I, Ravichandran M (2010) Estuarine , Coastal and Shelf Science Response of subsurface waters in the eastern Arabian Sea to tropical cyclones. Estuar Coast Shelf Sci 89:267–276. doi: 10.1016/j.ecss.2010.07.011

Rao R (1987) analysis on the thermal response of the upper Bay of Bengal to the forcing of pre-monsoon cyclonic storm and summer monsoonal onset during MONEX-. Mausam 11:1987

Rao RR, Sivakumar R (2003) Seasonal variability of sea surface salinity and salt budget of the mixed layer of the north Indian Ocean. J Geophys Res Ocean 108:. doi: 10.1029/2001jc000907

Rao VB, Ferreira CC, Franchito SH, Ramakrishna SSVS (2008) In a changing climate weakening tropical easterly jet induces more violent tropical storms over the north Indian Ocean. Geophys Res Lett 35:2–5. doi: 10.1029/2008GL034729

Ray, K., Giri, R. K., Ray, S. S., Dimri, A. P., & Rajeevan, M. (2021). An assessment of long-term changes in mortalities due to extreme weather events in India: A study of 50 years' data, 1970–2019. *Weather and Climate Extremes*, 32, 100315.

Reddy PRC, Salvekar PS, Nayak S (2008) Super cyclone induces a mesoscale phytoplankton bloom in the bay of bengal. IEEE Geosci Remote Sens Lett 5:588–592. doi: 10.1109/LGRS.2008.2000650

Reppin J, Schott FA, Fischer J, Quadfasel D (1999) Equatorial currents and transports in the upper central Indian Ocean: Annual cycle and interannual variability. J Geophys Res Ocean 104:15495–15514. doi: 10.1029/1999jc900093

Reul N, Quilfen Y, Chapron B, et al (2014) Multisensor observations of the Amazon-Orinoco river

plume interactions with hurricanes. J Geophys Res Ocean 119:8271–8295. doi: 10.1002/2014JC010107

Reynolds RW, Smith TM, Liu C, et al (2007) Daily high-resolution-blended analyses for sea surface temperature. J Clim 20:5473–5496. doi: 10.1175/2007JCLI1824.1

Roman-Stork HL, Subrahmanyam B (2020) The Impact of the Madden–Julian Oscillation on Cyclone Amphan (2020) and Southwest Monsoon Onset. Remote Sens 12:3011. doi: 10.3390/rs12183011

Roxy MK, Dasgupta P, McPhaden MJ, et al (2019) Twofold expansion of the Indo-Pacific warm pool warps the MJO life cycle. Nature 575:647–651. doi: 10.1038/s41586-019-1764-4

Roxy MK, Ritika K, Terray P, et al (2015) Drying of Indian subcontinent by rapid Indian ocean warming and a weakening land-sea thermal gradient. Nat Commun 6:1–10. doi: 10.1038/ncomms8423

Ryther JH, Hall JR, Pease AK, et al (1966) PRIMARY ORGANIC PRODUCTION IN RELATION TO THE CHEMSITRY OF THE WESTERN INDIAN OCEAN. Limnol Oceanogr 11:371–380

Sadhuram Y (2004) Record decrease of sea surface temperature following the passage of a super cyclone over the Bay of Bengal. Curr Sci 86:383–384

Sadhuram Y, Maneesha K, Murty TVR (2012) Intensification of Aila (May 2009) due to a warm core eddy in the north Bay of Bengal. Nat Hazards 63:1515–1525. doi: 10.1007/s11069-011-9837-1

Sadhuram Y, Maneesha K, Ramana Murty T V. (2010) Importance of upper ocean heat content in the intensification and translation speed of cyclones over the Bay of Bengal. Curr Sci 99:1191–1194

Sadhuram Y, Rao BP, Rao DP, et al (2004) Seasonal variability of cyclone heat potential in the Bay of Bengal. Nat Hazards 32:191–209. doi: 10.1023/B:NHAZ.0000031313.43492.a8

Sahoo B, Bhaskaran PK (2016) Assessment on historical cyclone tracks in the Bay of Bengal, east coast of India. Int J Climatol 36:95–109. doi: 10.1002/joc.4331

Saji N, Goswami B, Vinayachandran P, Yamagata T (1999) A dipole mode in the Tropical Ocean. Nature 401:360–363

Saji PK, Shenoi SC, Almeida A, Rao G (2000) Inertial currents in the Indian Ocean derived from satellite tracked surface drifters. Oceanol Acta 23:635–640. doi: 10.1016/S0399-1784(00)01108-7

Samson G, Giordani H, Caniaux G, Roux F (2009) Numerical investigation of an oceanic resonant regime induced by hurricane winds. Ocean Dyn 59:565–586. doi: 10.1007/s10236-009-0203-8

Sanabia ER, Barrett BS, Black PG, et al (2013) Real-time upper-ocean temperature observations from aircraft during operational hurricane reconnaissance missions: AXBT demonstration project year one results. Weather Forecast 28:1404–1422. doi: 10.1175/WAF-D-12-00107.1

Sanap SD, Mohapatra M, Ali MM, et al (2020) On the dynamics of cyclogenesis, rapid intensification and recurvature of the very severe cyclonic storm, Ockhi. J Earth Syst Sci 129:. doi: 10.1007/s12040-020-01457-2

Sandery PA, Brassington GB, Craig A, Pugh T (2010) Impacts of ocean-atmosphere coupling on tropical cyclone intensity change and ocean prediction in the Australian region. Mon Weather Rev 138:2074–2091. doi: 10.1175/2010MWR3101.1

Sarangi RK (2011) Impact of cyclones on the Bay of Bengal chlorophyll variability using remote sensing satellites. Indian J Mar Sci 40:794–801

Sarangi RK, Mishra MK, Chauhan P (2014) Remote Sensing Observations on Impact of Phailin Cyclone on Phytoplankton Distribution in Northern Bay of Bengal. IEEE J Sel Top Appl Earth Obs Remote Sens 8:539–549

Schade LR, Emanuel KA (1999) The ocean's effect on the intensity of tropical cyclones: Results from a simple coupled atmosphere-ocean model. J Atmos Sci 56:642–651. doi: 10.1175/1520-0469(1999)056<0642:TOSEOT>2.0.CO;2

Scoccimarro E, Fogli PG, Reed KA, et al (2017) Tropical cyclone interaction with the ocean: The role of high-frequency (subdaily) coupled processes. J Clim 30:145–162. doi: 10.1175/JCLI-D-16-0292.1

Sebastian M, Behera MR (2015) Impact of SST on tropical cyclones in North Indian Ocean. Procedia Eng 116:1072–1077. doi: 10.1016/j.proeng.2015.08.346

Sengupta D, Bharath Raj GN, Shenoi SSC (2006) Surface freshwater from Bay of Bengal runoff and Indonesian Throughflow in the tropical Indian Ocean. Geophys Res Lett 33:1–5. doi: 10.1029/2006GL027573

Sengupta D, Goddalehundi BR, Anitha DS (2008) Cyclone-induced mixing does not cool SST in the post-monsoon north Bay of Bengal. 6:1–6. doi: 10.1002/asl

Sharma N, Ali MM (2014) Importance of ocean heat content for cyclone studies. Oceanography 2:

Shay LK, Black PG, Mariano AJ, et al (1992) Upper ocean response to Hurricane Gilbert. J Geophys Res 97:. doi: 10.1029/92jc01586

Shay LK, Elsberry RL (1987) Near-inertial ocean current response to hurricane Frederic. J. Phys. Ocean. 17:1249–1269

Shay LK, Elsberry RL, Black PG (1989) Vertical Structure of the Ocean Current Response to a Hurricane. J Phys Oceanogr 19:649–669. doi: https://doi.org/10.1175/1520-0485(1989)019<0649:VSOTOC>2.0.CO;2

Shay LK, Goni GJ, Black PG (2000) Effects of a Warm Oceanic Feature on Hurricane Opal. Mon Weather Rev 128:1366–1383. doi: 10.1175/1520-0493(2000)128<1366:eoawof>2.0.co;2

Shen W, Ginis I (2003) Effects of surface heat flux-induced sea surface temperature changes on tropical cyclone intensity. Geophys Res Lett 30:1–4. doi: 10.1029/2003GL017878

Shengyan Y, Juncheng Z, Subrahmanyam MV (2019) Sea surface temperature cooling induced by

Tropical cyclone Hudhud over Bay of Bengal. Indian J Geo-Marine Sci 48:9–17

Shenoi SSC (2002) Differences in heat budgets of the near-surface Arabian Sea and Bay of Bengal: Implications for the summer monsoon. J Geophys Res 107:1–14. doi: 10.1029/2000jc000679

Shetye SR, Gouveia AD, Shankar D, et al (1996) Hydrography and circulation in the western Bay of Bengal during the northeast monsoon. J Geophys Res C Ocean 101:14011–14025. doi: 10.1029/95JC03307

Singh K, Panda J, Sahoo M, Mohapatra M (2019) Variability in Tropical Cyclone Climatology over North Indian Ocean during the Period 1891 to 2015. Asia-Pacific J Atmos Sci 55:269–287. doi: 10.1007/s13143-018-0069-0

Singh VK, Roxy MK, Deshpande M (2020) The unusual long track and rapid intensification of very severe cyclone Ockhi. Curr Sci 119:771–779

Singh VK, Roxy MK, Deshpande M (2021) Role of warm ocean conditions and the MJO in the genesis and intensification of extremely severe cyclone Fani. Sci Rep 11:3607. doi: 10.1038/s41598-021-82680-9

Sprintall J, Revelard A (2014) The Indonesian Throughflow response to Indo-Pacific climate variability. J Geophys Res Ocean 119:1161–1175. doi: 10.1002/2013JC009533.Received

Sprintall J, Tomczak M (1992) Evidence of the Barrier Layer in the Surface Layer of the Tropics ocean surface mixed layer generally denotes a quasi- kinetic energy and potential energy processes mentioned its degree state. J Geophys Res 97:7305–7316

Sreenivas P, Chowdary JS, Gnanaseelan C (2012) Impact of tropical cyclones on the intensity and phase propagation of fall Wyrtki jets. Geophys Res Lett 39:1–6. doi: 10.1029/2012GL053974

Sreenivas P, Gnanaseelan C (2014) Impact of oceanic processes on the life cycle of severe cyclonic storm "Jal." IEEE Geosci Remote Sens Lett 11:519–523. doi: 10.1109/LGRS.2013.2271512

Srinivas C V., Mohan GM, Naidu C V., et al (2016) Impact of air-sea coupling on the simulation of tropical cyclones in the North Indian Ocean using a simple 3-D ocean model coupled to ARW. J Geophys Res 121:9400–9421. doi: 10.1002/2015JD024431

Sriver RL, Huber M (2007) Observational evidence for an ocean heat pump induced by tropical cyclones. Nature 447:577–580. doi: 10.1038/nature05785

Sriver RL, Huber M, Nusbaumer J (2008) Investigating tropical cyclone-climate feedbacks using the TRMM Microwave Imager and the Quick Scatterometer. Geochemistry, Geophys Geosystems 9:. doi: 10.1029/2007GC001842

Subrahmanyam B, Murty VSN, Sharp RJ, O'Brien JJ (2005) Air-sea coupling during the tropical cyclones in the Indian Ocean: A case study using satellite observations. Pure Appl Geophys 162:1643–1672. doi: 10.1007/s00024-005-2687-6

Subrahmanyam B, Rao KH, Srinivasa Rao N, et al (2002) Influence of a tropical cyclone on chlorophyll-a concentration in the Arabian Sea. Geophys Res Lett 29:22-1-22–4. doi: 10.1029/2002gl015892

Sun C, Wang X, Cui X, et al (2015) Satellite derived upper ocean thermal structure and its application to tropical cyclone intensity forecasting in the Indian Ocean. Chinese J Oceanol Limnol 33:1219–1232. doi: 10.1007/s00343-015-4114-x

Sun L, Yang YJ, Xian T, et al (2010) Strong enhancement of chlorophyll a concentration by a weak typhoon. Mar Ecol Prog Ser 404:39–50. doi: 10.3354/meps08477

Sun Y, Zhong Z, Li T, et al (2017) Impact of Ocean Warming on Tropical Cyclone Size and Its Destructiveness. Sci Rep 7:1–10. doi: 10.1038/s41598-017-08533-6

Suneeta P, Sadhuram Y (2018) Tropical Cyclone Genesis Potential Index for Bay of Bengal During Peak Post-Monsoon (October-November) Season Including Atmosphere-Ocean Parameters. Mar Geod 41:86–97. doi: 10.1080/01490419.2017.1394404

Taylor KE, Stouffer RJ, Meehl GA (2012) An overview of CMIP5 and the experiment design. Bull Am Meteorol Soc 93:485–498. doi: 10.1175/BAMS-D-11-00094.1

Thadathil P, Muraleedharan PM, Rao RR, et al (2007) Observed seasonal variability of barrier layer in the Bay of Bengal. J Geophys Res Ocean 112:. doi: 10.1029/2006JC003651

Thakur MK, Kumar TVL, Dwivedi S (2018) On the rainfall asymmetry and distribution in tropical cyclones over Bay of Bengal using TMPA and GPM rainfall products. Nat Hazards 0123456789: doi: 10.1007/s11069-018-3426-5

Trenberth KE, Cheng L, Jacobs P, et al (2018) Hurricane Harvey Links to Ocean Heat Content and Climate Change Adaptation. Earth's Futur 6:730–744. doi: 10.1029/2018EF000825

Tsuboi A, Takemi T (2014) The interannual relationship between MJO activity and tropical cyclone genesis in the Indian Ocean. Geosci Lett 1:9. doi: 10.1186/2196-4092-1-9

Tummala SK, Mupparthy RS, Masuluri NK, Nayak S (2009) Phytoplankton bloom due to Cyclone Sidr in the central Bay of Bengal. J Appl Remote Sens 3:

Uddin MJ, Nasrin ZM, Li Y (2021) Effects of vertical wind shear and storm motion on tropical cyclone rainfall asymmetries over the North Indian Ocean. Dyn Atmos Ocean 93:101196. doi: 10.1016/j.dynatmoce.2020.101196

Venkatesan R, Joseph KJ, Prasad CA, et al (2020) Differential upper ocean response depicted in moored buoy observations during the pre-monsoon cyclone Viyaru. Curr Sci 118:1760–1767. doi: 10.18520/cs/v118/i11/1760-1767

Venkatesan R, Mathew S, Vimala J, et al (2014) Signatures of very severe cyclonic storm Phailin in met-ocean parameters by moored buoy network in the Bay of Bengal. Curr Sci 107:589–595

Vidya PJ, Balaji M, Murali RM (2021) Cyclone Hudhud-eddy induced phytoplankton bloom in the northern Bay of Bengal using a coupled model. Prog Oceanogr 197:102631. doi: 10.1016/j.pocean.2021.102631

Vidya PJ, Das S, Murali.R M (2017) Contrasting Chl-a responses to the tropical cyclones Thane and Phailin in the Bay of Bengal. J Mar Syst 165:103–114. doi: 10.1016/j.jmarsys.2016.10.001

Vidya PJ, Ravichandran M, Murtugudde R, et al (2020) Increased cyclone destruction potential in the Southern Indian Ocean. Environ Res Lett 16:. doi: 10.1088/1748-9326/abceed

Vinayachandran PN, Kagimoto T, Masumoto Y, et al (2005) Bifurcation of the East India Coastal Current east of Sri Lanka. Geophys Res Lett 32:6–9. doi: 10.1029/2005GL022864

Vinayachandran PN, Mathew S (2003) Phytoplankton bloom in the Bay of Bengal during the northeast monsoon and its intensification by cyclones. Geophys Res Lett 30:1999–2002. doi: 10.1029/2002GL016717

Vinayachandran PN, Murty VSN, Babu VR (2002) Observations of barrier layer formation in the Bay of Bengal during summer monsoon. J Geophys Res Ocean 107:1–9. doi: 10.1029/2001jc000831

Vincent EM, Lengaigne M, Madec G, et al (2012a) Processes setting the characteristics of sea surface cooling induced by tropical cyclones. J Geophys Res Ocean 117:1–18. doi: 10.1029/2011JC007396

Vincent EM, Lengaigne M, Vialard J, et al (2012b) Assessing the oceanic control on the amplitude of sea surface cooling induced by tropical cyclones. J Geophys Res Ocean 117:. doi: 10.1029/2011JC007705

Vinod KK, Soumya M, Tkalich P, Vethamony P (2014) Ocean - Atmosphere interaction during thane cyclone: A numerical study using WRF. Indian J Geo-Marine Sci 43:1230–1235

Vinodhkumar B, Busireddy NKR, Ankur K, et al (2021) On Occurrence of Rapid Intensification and Rainfall changes in Tropical Cyclones over the North Indian Ocean. Int J Climatol 1–13. doi: 10.1002/joc.7268

Vissa NK, Satyanarayana ANV, Prasad Kumar B (2013a) Intensity of tropical cyclones during pre- and post-monsoon seasons in relation to accumulated tropical cyclone heat potential over Bay of Bengal. Nat Hazards 68:351–371. doi: 10.1007/s11069-013-0625-y

Vissa NK, Satyanarayana ANV, Prasad Kumar B (2012) Response of upper ocean during passage of mala cyclone utilizing argo data. Int J Appl Earth Obs Geoinf 14:149–159. doi: 10.1016/j.jag.2011.08.015

Vissa NK, Satyanarayana AN V, Kumar BP (2013b) Response of Upper Ocean and Impact of Barrier Layer on Sidr Cyclone Induced Sea Surface Cooling. 48:279–288

Wada A, Usui N (2007) Importance of tropical cyclone heat potential for tropical cyclone intensity and intensification in the Western North Pacific. J Oceanogr 63:427–447. doi: 10.1007/s10872-007-0039-0

Walker ND, Leben RR, Pilley CT, et al (2014) Slow translation speed causes rapid collapse of northeast Pacific Hurricane Kenneth over cold core eddy. Geophys Res Lett 41:7595–7601. doi: 10.1002/2014GL061584

Walsh KJE, Mcbride JL, Klotzbach PJ, et al (2016) Tropical cyclones and climate change. Wiley Interdiscip Rev Clim Chang 7:65–89. doi: 10.1002/wcc.371

Wang D, Zhao H (2008) Estimation of phytoplankton responses to Hurricane Gonu over the Arabian Sea based on ocean color data. Sensors 8:4878–4893. doi: 10.3390/s8084878

Wang JW, Han W (2014) The Bay of Bengal upper-ocean response to tropical cyclone forcing during 1999. J Geophys Res Ocean 119:98–120. doi: 10.1002/2013JC008965

Wang JW, Han W, Sriver RL (2012a) Impact of tropical cyclones on the ocean heat budget in the Bay of Bengal during 1999: 2. Processes and interpretations. J Geophys Res Ocean 117:9021. doi: 10.1029/2012JC008373

Wang XD, Han GJ, Qi YQ, Li W (2011) Impact of barrier layer on typhoon-induced sea surface cooling. Dyn Atmos Ocean 52:367–385. doi: 10.1016/j.dynatmoce.2011.05.002

Wang Z, DiMarco SF, Stössel MM, et al (2012b) Oscillation responses to tropical Cyclone Gonu in northern Arabian Sea from a moored observing system. Deep Res Part I Oceanogr Res Pap 64:129–145. doi: 10.1016/j.dsr.2012.02.005

Webster PJ (2013) Meteorology: Improve weather forecasts for the developing world. Nature 493:17–19. doi: 10.1038/493017a

Wyrtki K (1973) An equatorial jet in the Indian Ocean. Science (80- ) 181:262–264. doi: 10.1126/science.181.4096.262

Xu W, Jiang H, Kang X (2014) Rainfall asymmetries of tropical cyclones prior to, during, and after making landfall in South China and Southeast United States. Atmos Res 139:18–26. doi: 10.1016/j.atmosres.2013.12.015

Yablonsky RM, Ginis I, Thomas B, et al (2015) Description and analysis of the ocean component of NOAA'S operational hurricane weather research and forecasting model (HWRF). J Atmos Ocean Technol 32:144–163. doi: 10.1175/JTECH-D-14-00063.1

Ye HJ, Kalhoro MA, Sun J, Tang D (2018) Chlorophyll blooms induced by tropical cyclone vardah in the Bay of Bengal. Indian J Geo-Marine Sci 47:1383–1390

Yokoi S (2010) Environmental and External Factors in the Genesis of Tropical Cyclone Nargis in April 2008 over the Bay of Bengal. 88:425–435. doi: 10.2151/jmsj.2010-310

Yu L (2019) Global air-sea fluxes of heat, fresh water, and momentum: Energy budget closure and unanswered questions. Ann Rev Mar Sci 11:227–248. doi: 10.1146/annurev-marine-010816-060704

Yu L, McPhaden MJ (2011) Ocean preconditioning of Cyclone Nargis in the Bay of Bengal: Interaction between Rossby waves, Surface Fresh Waters, and Sea Surface Temperatures. J Phys Oceanogr 41:1741–1755. doi: 10.1175/2011JPO4437.1

Yu Z, Wang Y, Xu H (2015) Observed rainfall asymmetry in tropical cyclones making landfall over China. J Appl Meteorol Climatol 54:117–136. doi: 10.1175/JAMC-D-13-0359.1

Yuan JP, Cao J (2013) North Indian Ocean tropical cyclone activities influenced by the Indian Ocean Dipole mode. Sci China Earth Sci 56:855–865. doi: 10.1007/s11430-012-4559-0

Zhang C (1993) Large-Scale Variability of Atmospheric Deep Convection in Relation to Sea Surface Temperature in the Tropics. J Clim 6:1898–1913

Zhang H, Wu R, Chen D, et al (2018) Net Modulation of Upper Ocean Thermal Structure by Typhoon Kalmaegi (2014). J Geophys Res Ocean 123:7154–7171. doi:

10.1029/2018JC014119

Zheng ZW, Ho CR, Kuo NJ (2008) Importance of pre-existing oceanic conditions to upper ocean response induced by Super Typhoon Hai-Tang. Geophys Res Lett 35:1–5. doi: 10.1029/2008GL035524

Zhu H, Ulrich W, Smith RK (2004) Ocean effects on tropical cyclone intensification and inner-core asymmetries. J Atmos Sci 61:1245–1258. doi: 10.1175/1520-0469(2004)061<1245:OEOTCI>2.0.CO;2

Zhu T, Zhang D-L (2006) The impact of storm-induced SST cooling on hurricane intensity. Adv Atmos Sci 23:14–22

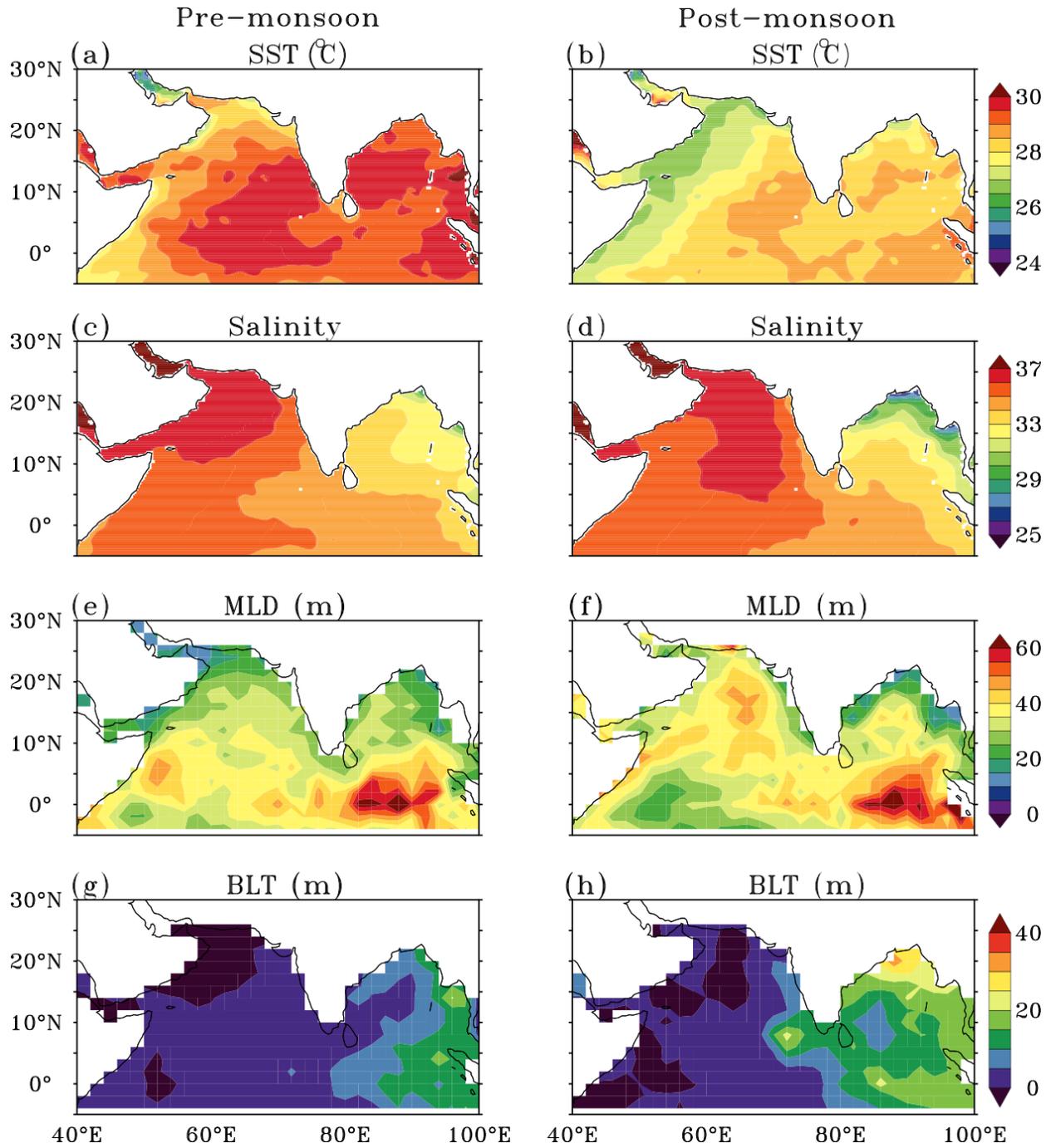

**Figure 1**. Climatology of (a-b) SSTs (°C) (c-d) sea surface salinity (e-f) mixed layer depth (MLD, m) and (g-h) barrier layer thickness (BLT, m). The left panel is for the pre-monsoon (April-June) season and right panel is for the post-monsoon (October-December) season. The climatology of the SST and salinity is obtained from the World Ocean Atlas 2018. The climatology of mixed layer depth and barrier layer thickness is obtained from de Boyer climatology of the mixed layer and barrier layer (de Boyer Montégut et al. 2007) where the mixed layer depth is defined as the depth at which the change in the density is 0.03 kg m$^{-3}$ as compared to the density at the surface.

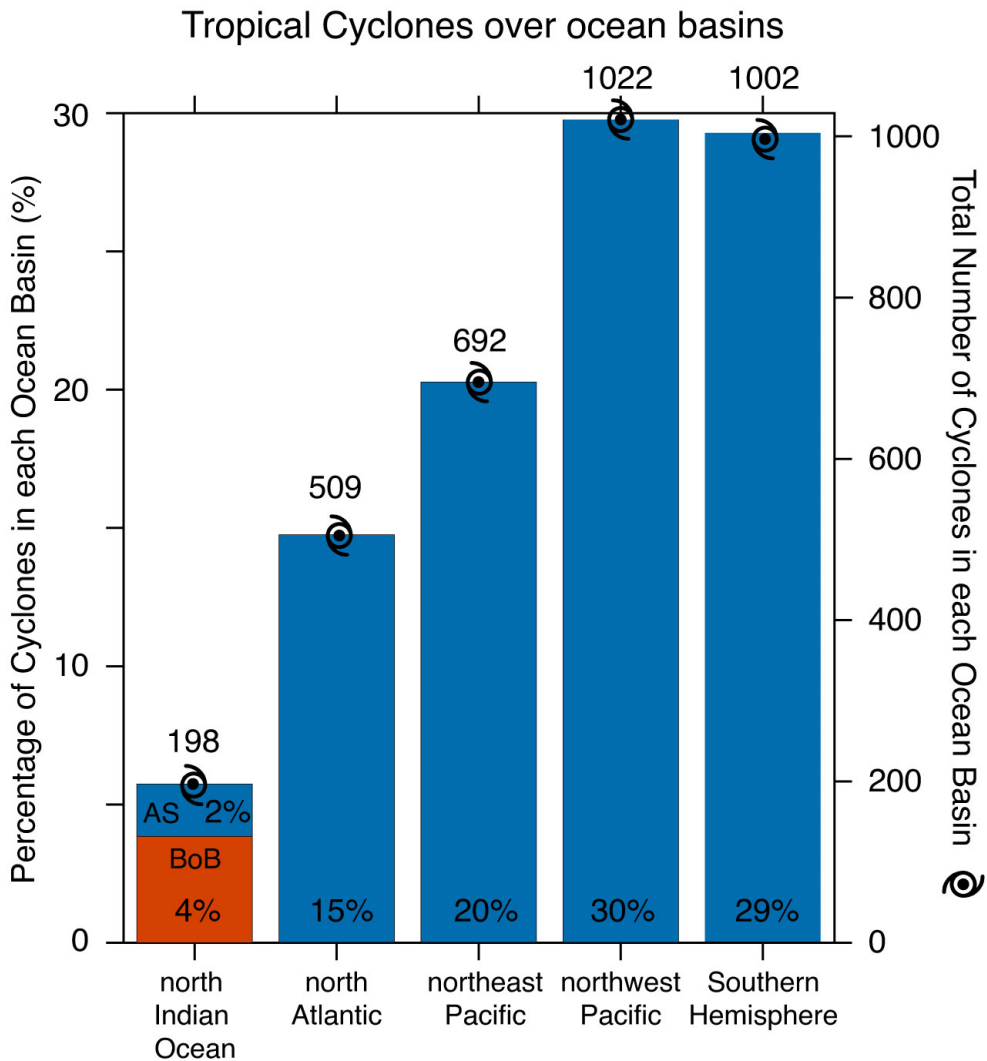

**Figure 2**. Percentage and the total number of cyclones in each ocean basins, during the period 1980-2019. 4% of total cyclones occur in the Bay of Bengal and less than 2% occur in the Arabian Sea. The total number of cyclones in the north Indian Ocean is less than 6% only, but accounts for more than 80% of the global fatalities due to cyclones. The cyclone data is obtained from the IBTrACS dataset.

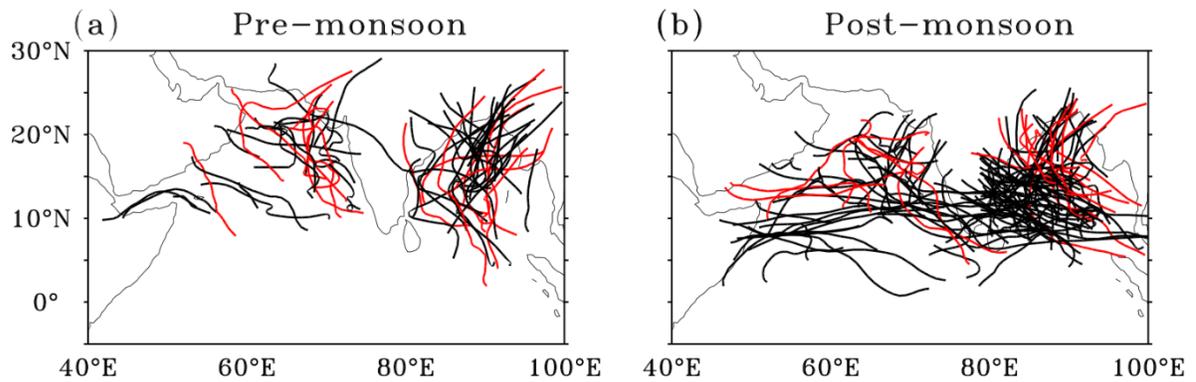

**Figure 3**. Track of the cyclones in the Arabian Sea and the Bay of Bengal during the period 1980–2019 in (a) pre-monsoon and (b) post-monsoon seasons. Black color tracks are for cyclones with maximum wind speed ≤ 95 knots (Category 2 or lower). Red color tracks are for cyclones with maximum wind speed ≥ 100 knots (Category 3 or higher).

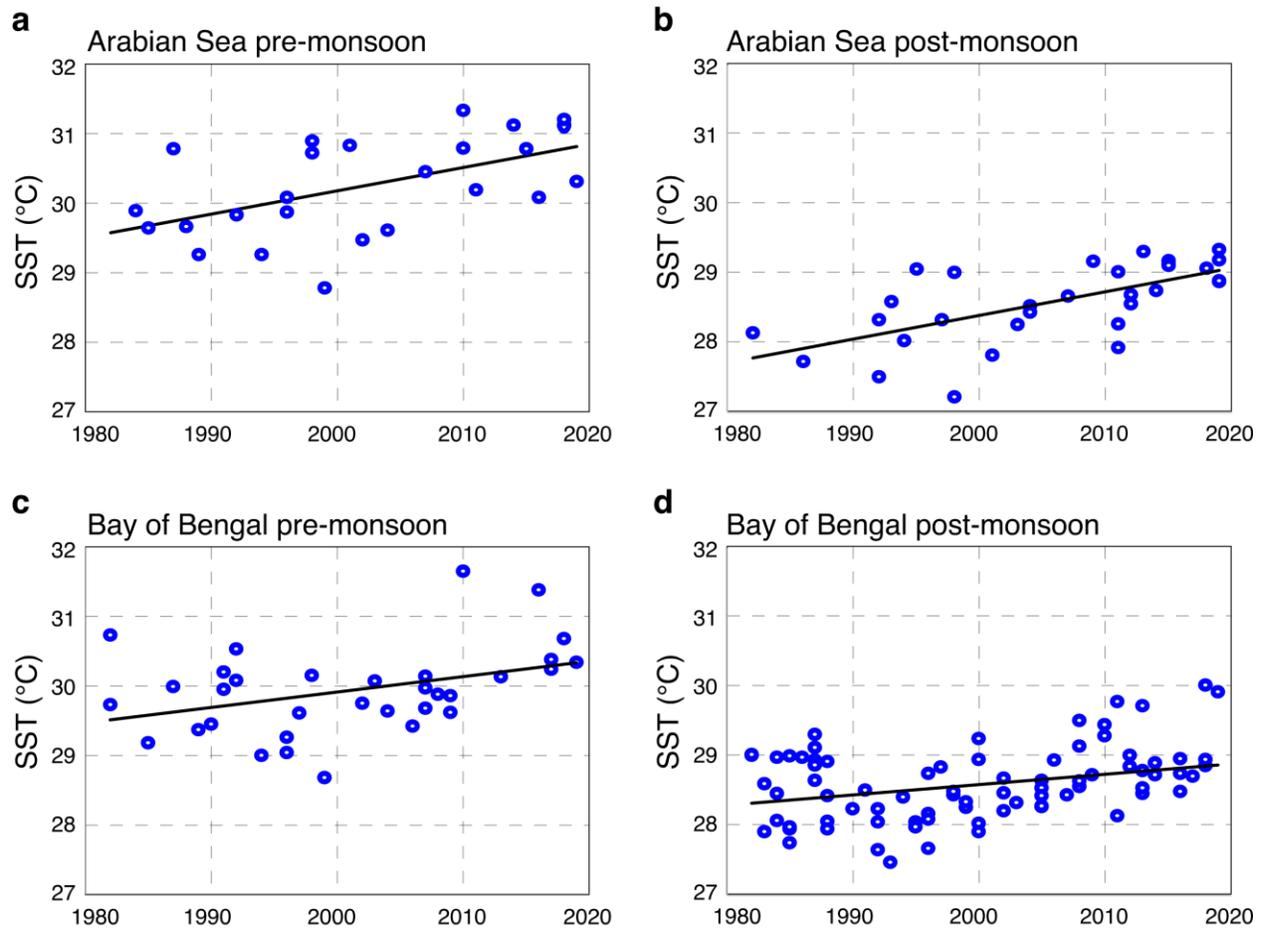

**Figure 4**. SSTs averaged for a week (day -7 to day -1) prior to the day of cyclogenesis (day 0), over a 5°x5° region around the genesis center—for (a) Arabian Sea pre-monsoon, (b) Arabian Sea post-monsoon, (c) Bay of Bengal pre-monsoon and (d) Bay of Bengal post-monsoon. The total SST change (for SSTs prior to cyclones) over the Arabian Sea during 1982–2019 is 1.4°C in the pre-monsoon and 1.2°C in the post-monsoon season. Over the Bay of Bengal, the observed change in SST is 0.8°C in the pre-monsoon and 0.5°C in the post-monsoon. The SST data is obtained from the Optimum Interpolation Sea Surface Temperature (OISST) provided by the National Oceanic and Atmospheric Administration (NOAA).

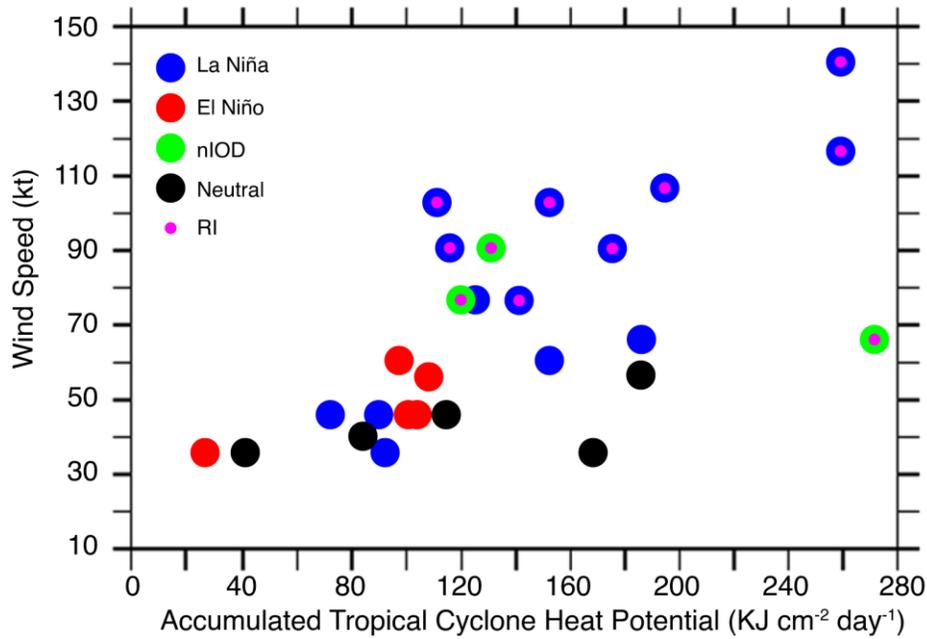

**Figure 5**. Scatter plot of maximum wind speed (knots) and accumulated tropical cyclone heat potential (kJ cm$^{-2}$/day) during the period 1993–2011, in the Bay of Bengal. Cyclones formed during La Niña, El Niño, negative Indian Ocean Dipole (nIOD) and neutral years are marked in blue, red, green and black circles, respectively. Cyclones experiencing rapid intensification under La Niña and nIOD conditions are marked in pink color. Redrawn based on Girishkumar et al. (2015).

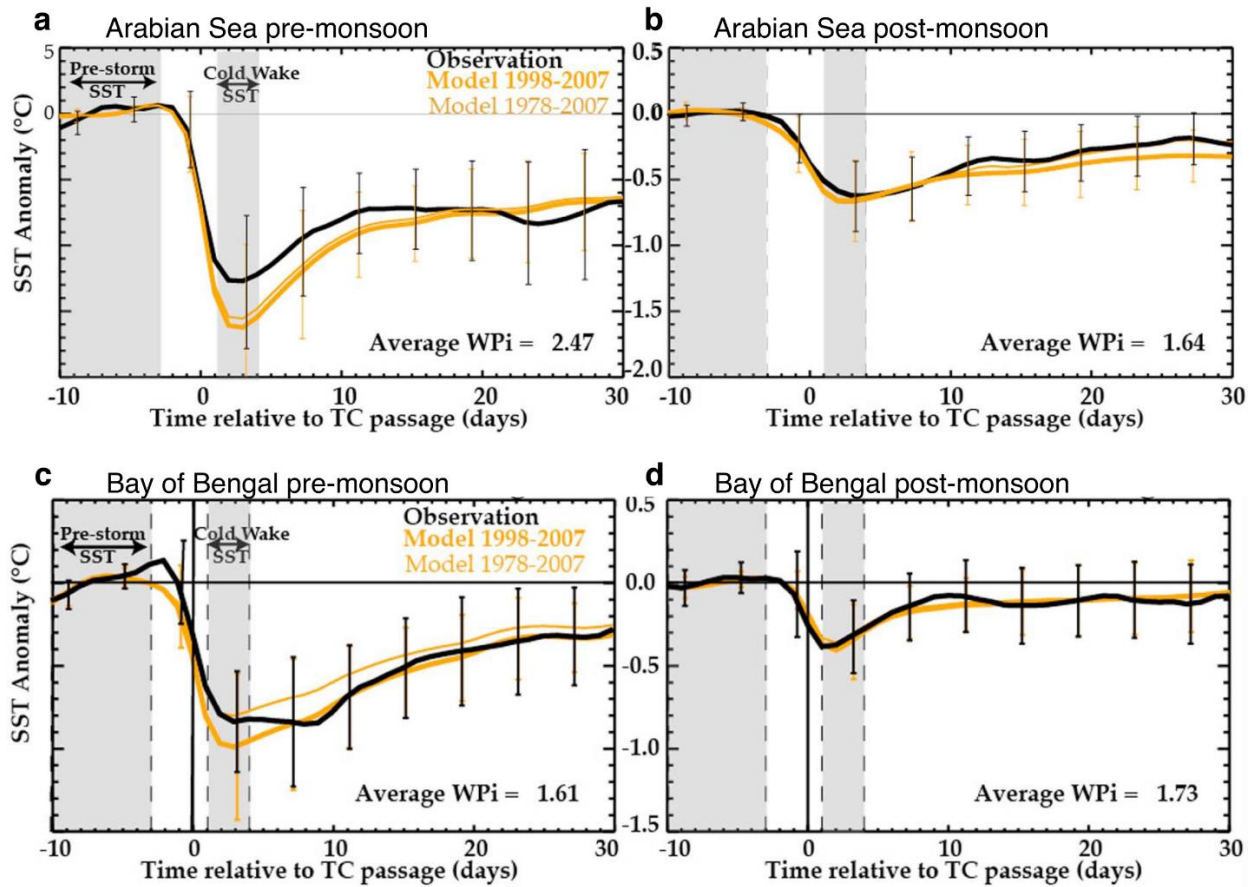

**Figure 6.** Composite evolution of cyclone induced SST cooling (°C) within 200 km of cyclone tracks in the Arabian Sea and Bay of Bengal during (a, c) pre-monsoon and (b, d) post-monsoon seasons for observations (black line) and the model over the 1998–2007 period (thick orange line) and the model over the 1978–2007 period (thin orange line). Vertical bars (black for the observations and orange for the model) indicate the spread around the mean, evaluated from the lower and upper quartiles. Adapted from Neetu et al. (2012).

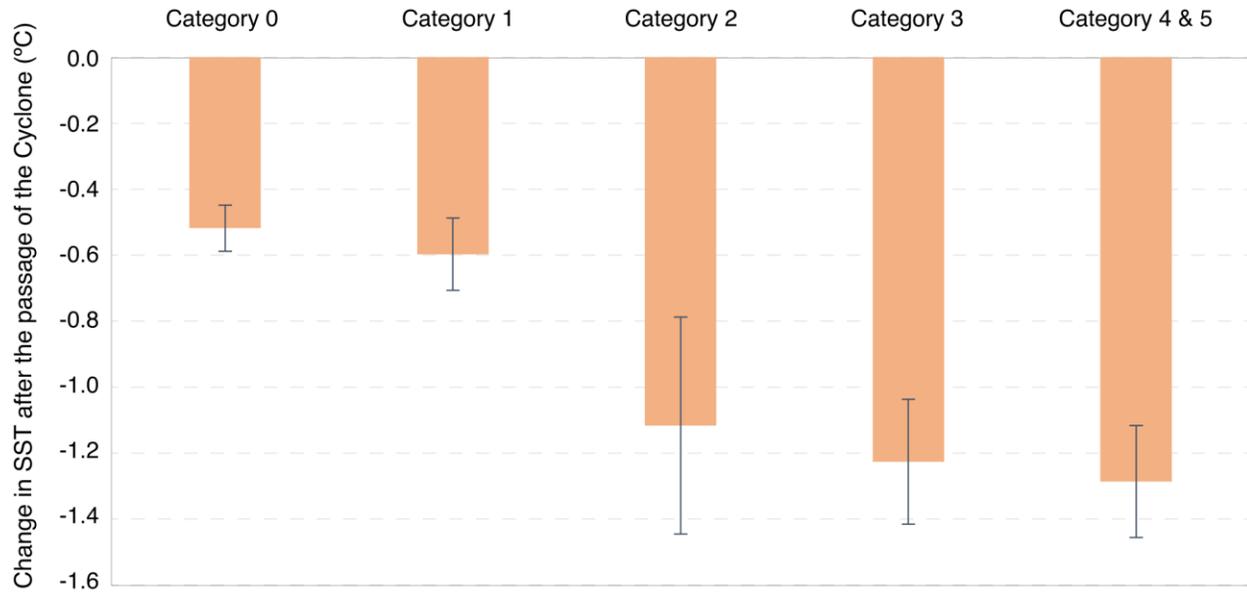

**Figure 7**. Composite change in SST due to cyclones in the north Indian Ocean during the period 1982–2019, according to the cyclone categories. Based on the maximum sustained wind speed, the cyclones are categorized as category 1 (119–153 km h$^{-1}$), category 2 (154–177 km h$^{-1}$), category 3 (178–208 km h$^{-1}$), category 4 (209–251 km h$^{-1}$) and category 5 ($\geq$ 252 km h$^{-1}$). Tropical storms (JTWC definition, 63–118 km h$^{-1}$) that are considered as cyclones by IMD are included in Category 0. The change in SST is estimated as the difference between SST averaged from day +1 to day +5 after the passage of the cyclone passage and the SST averaged from day -6 to day -2 before the passage of the cyclone, over a 1°x1° region where and when the cyclone attained its maximum wind speed (referred to as day 0). The whiskers on the bars denote the standard error for each cyclone category, estimated by dividing the standard deviation with the square root of the number of observations. The cyclone data is obtained from IBTrACS and the SST data is obtained from OISST provided by NOAA.

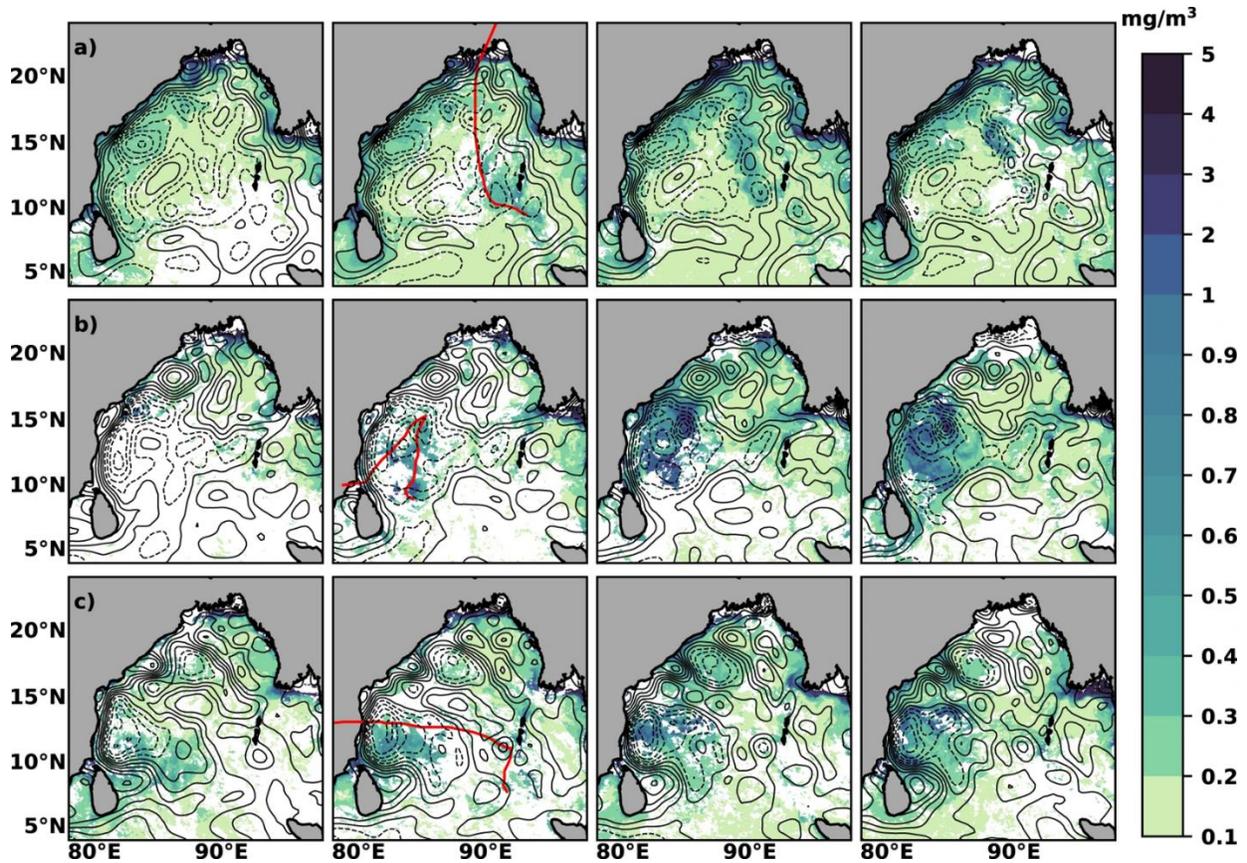

**Figure 8**. The average chlorophyll, overlaid with sea surface height anomalies contours (solid is positive and dashed is negative), for 5 days before cyclone, during the entire cyclone period, five days immediately after the passage of cyclone and the next 5 days—for cyclones (a) Sidr (2007), (b) Madi (2013), and (c) Vardah (2016). The tracks of each cyclone are shown by red color. Adapted from (Kuttippurath et al. 2021).

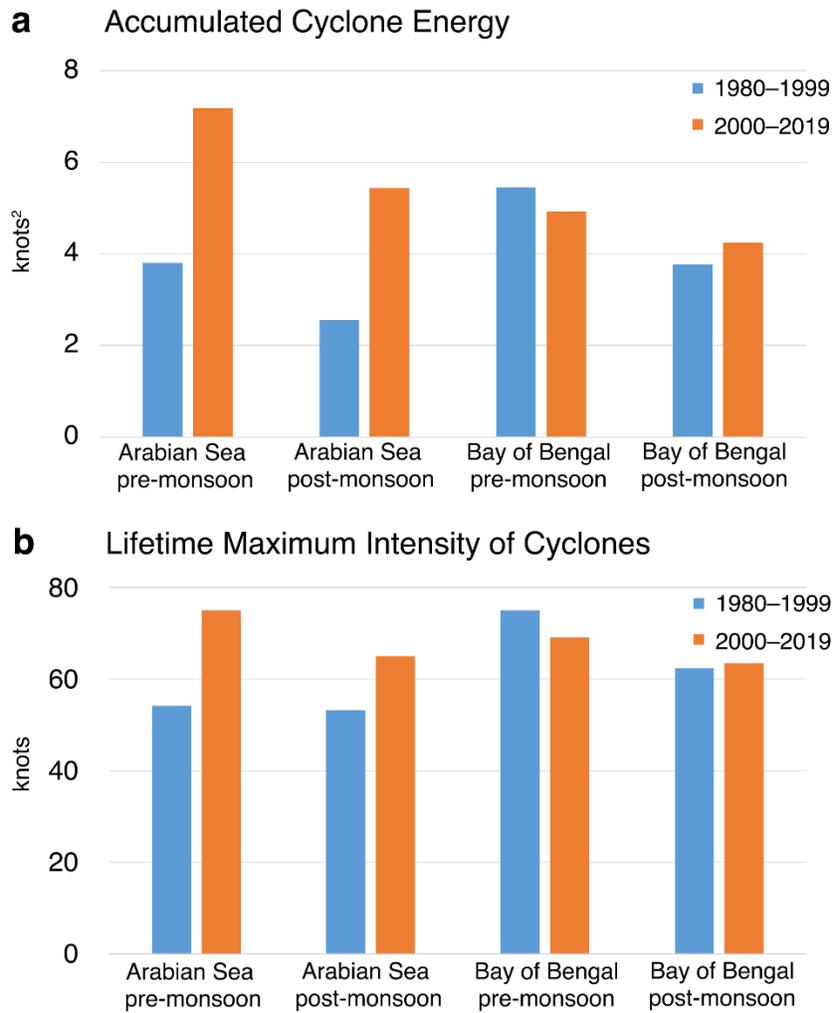

**Figure 9**. (a) Accumulated cyclone energy (knots$^2$) and (b) lifetime maximum intensity (knots) of cyclones during the period 1980–1999 (blue bars) and 2000–2019 (orange bars) in the Arabian Sea and the Bay of Bengal basin during the pre-monsoon and post-monsoon seasons.

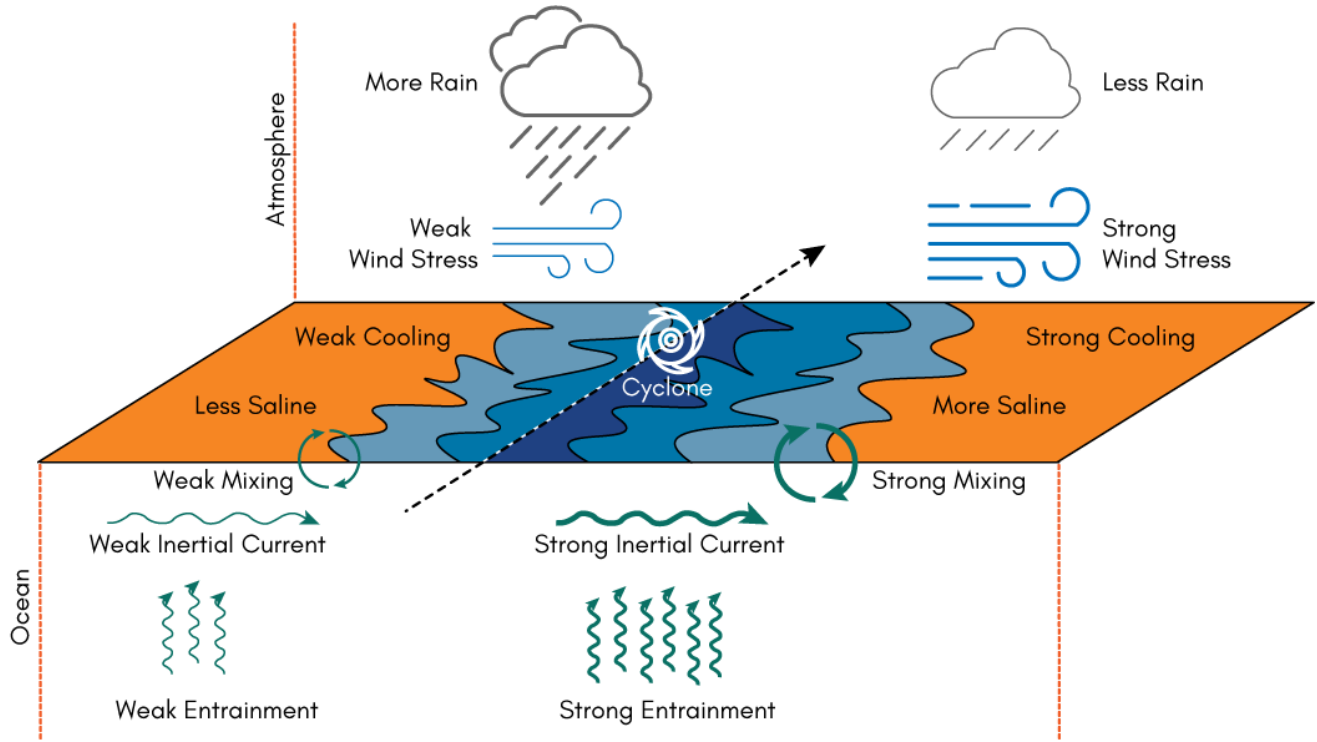

**Figure 10**. Schematic highlighting some of the key open-ocean processes involved in the asymmetric response of the ocean-cyclone interactions over the north Indian Ocean.

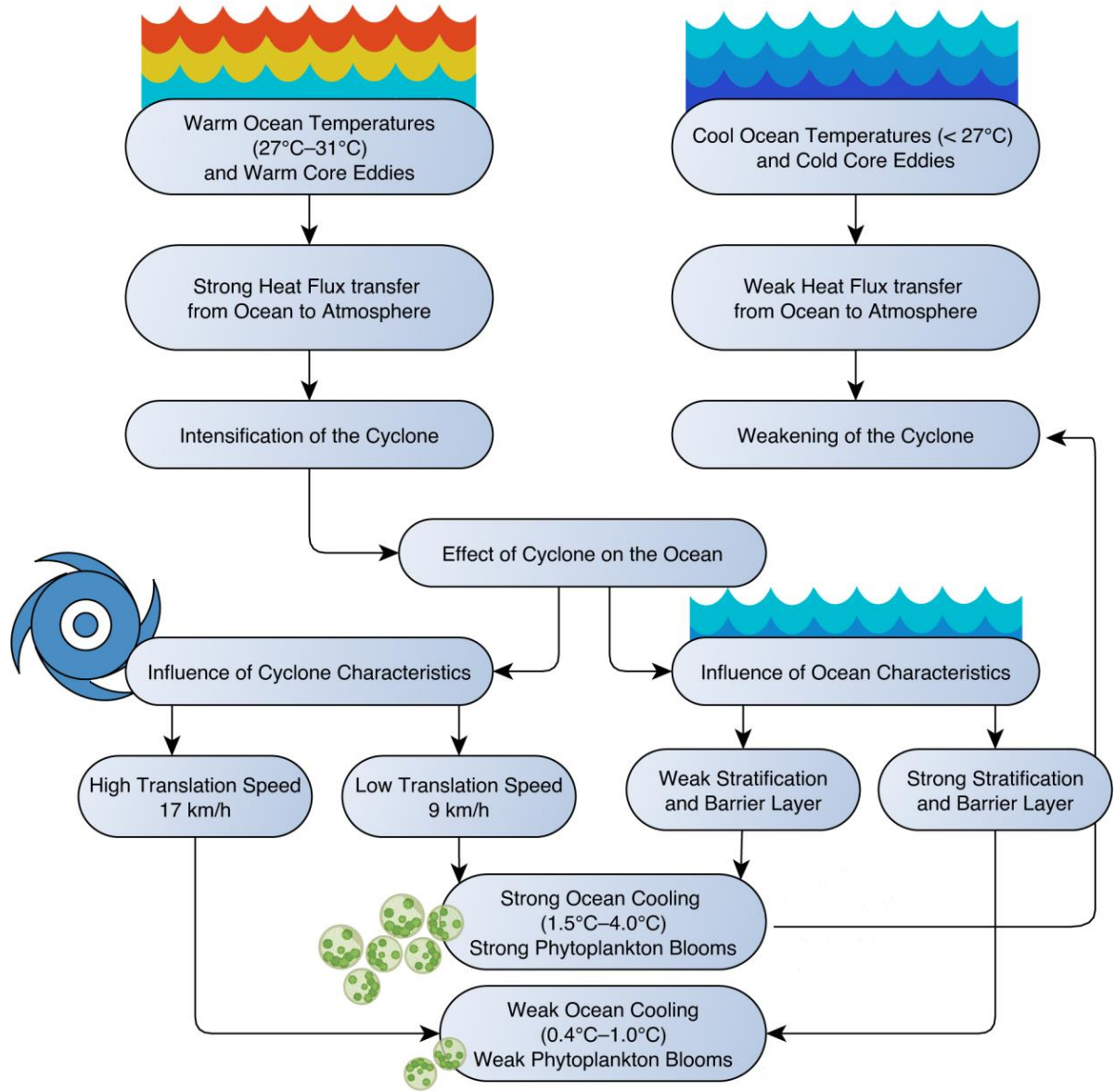

**Figure 11**. Schematic highlighting some of the key processes involved in the ocean-cyclone interactions in the north Indian Ocean. Quantification of the ocean temperatures, translation speed, and ocean cooling is based on the studies discussed in this review.